\documentclass[aps,twocolumn,prb,floatfix,showpacs,superscriptaddress,amsmath,amssymb,floatfix]{revtex4-1}

\usepackage{graphicx}
\usepackage{dcolumn}
\usepackage{bm}
\usepackage{color}
\usepackage{url}
\usepackage{tabularx}
\usepackage{array}
\usepackage{booktabs}
\usepackage{comment}
\usepackage{footnote}
\usepackage[normalem]{ulem}
\usepackage{multirow}
\usepackage{dcolumn}
\usepackage{units}
\usepackage[colorlinks]{hyperref}
	
\usepackage[T1]{fontenc}
\usepackage[utf8]{inputenc}
\usepackage[english]{babel}
	
\usepackage{johansmacros}
	
\usepackage[cmyk,dvipsnames]{xcolor}
	
\usepackage{braket}

\usepackage{algorithm}

\usepackage{algpseudocode}

\definecolor{pacificb}{HTML}{1CA9C9}
\definecolor{magenta}{HTML}{FF00FF}
\definecolor{goldbrown}{HTML}{994C00}
\definecolor{darkblue}{HTML}{0000CC}
\definecolor{ocre}{HTML}{923232}

\newcommand{\anna}[1]{\textsf{\textcolor{pacificb}{\textit{#1}}}}

\newcommand{\mat}[1]{\boldsymbol{\mathsf{#1}}}

\renewcommand{\vec}[1]{\boldsymbol{#1}}

\newcommand{\angstrom}{\mbox{\normalfont\AA}}
\begin{document}

\title{General method for atomistic spin-lattice dynamics with first principles accuracy}

\author{Johan Hellsvik}
\email{hellsvik@kth.se}
\affiliation{Nordita, Roslagstullsbacken 23, SE-106 91 Stockholm, Sweden}
\affiliation{Department of Physics, KTH Royal Institute of Technology, SE-106 91 Stockholm, Sweden}

\author{Danny Thonig}
\affiliation{Department of Physics and Astronomy, Materials Theory Division, Uppsala University, Box 516, SE-75120 Uppsala, Sweden}

\author{Klas Modin}
\affiliation{Department of Mathematics, Chalmers University of Technology, Gothenburg, Sweden}
\affiliation{Department of Mathematics, University of Gothenburg, Gothenburg, Sweden}

\author{Diana Iu\c{s}an}
\affiliation{Department of Physics and Astronomy, Materials Theory Division, Uppsala University, Box 516, SE-75120 Uppsala, Sweden}

\author{Anders Bergman}
\affiliation{Department of Physics and Astronomy, Materials Theory Division, Uppsala University, Box 516, SE-75120 Uppsala, Sweden}

\author{Olle Eriksson}
\affiliation{Department of Physics and Astronomy, Materials Theory Division, Uppsala University, Box 516, SE-75120 Uppsala, Sweden}
\affiliation{School of Science and Technology, \"Orebro University, SE-701 82 \"Orebro, Sweden}

\author{Lars Bergqvist}
\affiliation{Department of Applied Physics, School of Engineering Sciences, KTH Royal Institute of Technology, Electrum 229, SE-16440 Kista, Sweden}
\affiliation{SeRC (Swedish e-Science Research Center), KTH Royal Institute of Technology, SE-10044 Stockholm, Sweden}

\author{Anna Delin}
\email{annadel@kth.se}
\affiliation{Department of Applied Physics, School of Engineering Sciences, KTH Royal Institute of Technology, Electrum 229, SE-16440 Kista, Sweden}
\affiliation{SeRC (Swedish e-Science Research Center), KTH Royal Institute of Technology, SE-10044 Stockholm, Sweden}
\affiliation{Department of Physics and Astronomy, Materials Theory Division, Uppsala University, Box 516, SE-75120 Uppsala, Sweden}

\date{\today}


\begin{abstract}
We present a computationally efficient general first-principles based method for spin-lattice simulations for solids and clusters. The method is based on a coupling of atomistic spin dynamics and molecular dynamics simulations, expressed through a spin-lattice Hamiltonian, where the bilinear magnetic term is expanded up to second order in displacement. The effect of first order spin-lattice coupling on the magnon and phonon dispersion in bcc Fe is reported as an example, and we observe good agreement with previous simulations. In addition, we also illustrate the coupled spin-lattice dynamics method on a more conceptual level, by exploring dissipation-free spin and lattice motion of small magnetic clusters (a dimer, trimer and quadmer). The here discussed method opens the door for a quantitative description and understanding of the microscopic origin of many fundamental phenomena of contemporary interest, such as ultrafast demagnetization, magnetocalorics, and spincaloritronics. \end{abstract}

\maketitle

\section{Introduction}\label{sec:intro}

The way in which atoms vibrate around their equilibrium positions as a function of temperature is of fundamental importance for a range of physical properties of solids, for example thermal expansion, specific heat, thermal conductivity, and superconductivity. These vibrations can be studied computationally using molecular dynamics (MD) simulations, which is nowadays a mature and widely used technique in computational materials science. Phonon spectra and other properties related to the atomic vibrations are  nowadays routinely computed. To address systems with millions of atoms with MD, empirical potentials are usually necessary. Only for relatively small systems are MD simulations at the first-principles level feasible.\cite{Car1985,Marx2009aim}

In systems with magnetic order, there also exist collective motion of the spins, in addition to the above mentioned lattice vibrations. The standard approach to simulate the time evolution of the spin texture is to propagate the Landau-Lishitz-Gilbert (LLG) equation. Both continuum models (usually called "micromagnetics") \cite{Fidler2000} and atomistic models, so-called atomistic spin dynamics (ASD),\cite{Antropov1996,Nowak2005,Skubic2008,Evans2014,Eriksson2017asd} have been developed. In principle, spin motion can also be addressed directly at the first-principles level,\cite{Antropov1996,Katsnelson2003,Bhattacharjee2012,Fransson2017} as is possible in the framework of time-dependent density functional theory,\cite{Krieger2015} although this normally requires too much computing resources and time to be a realistic approach for most systems of interest. Also, in order to take dissipation and fluctuations into account in spin-dynamic simulations, a phenomenological stochastic approach is normally employed, for details see e.g. Ref. \onlinecite{Eriksson2017asd}. A full microscopic description of dissipation would require explicit descriptions of all the spin-electron couplings as well as all spin-lattice couplings.

In reality, the atomic magnetic moment and lattice degrees of freedom are always more or less coupled, a coupling which is mediated by the electronic subsystem. These couplings will determine, for instance, how fast it is possible to change the magnetic state of a material, and how relaxation of phonon and electronic subsystems proceed after excitation with ultra-short laser pulses.\cite{Maldonado2017} In addition, these couplings may shine light on angular momentum transfer between the spin and lattice subsystems in pump-probe experiments. 

Already in the original work on the equations of motion of atomic moments\cite{Antropov1996} a formalism that allows for coupled spin-lattice simulations was provided. Recently, such coupling was derived from a different ansatz, in which the role of the underlying electronic structure in mediating these couplings is explicitly considered.\cite{Fransson2017} However, the application to real materials remains a challenge for any formulation of spin-lattice simulations.

\begin{figure}[ht]
\centering
\includegraphics[width=0.50\columnwidth]{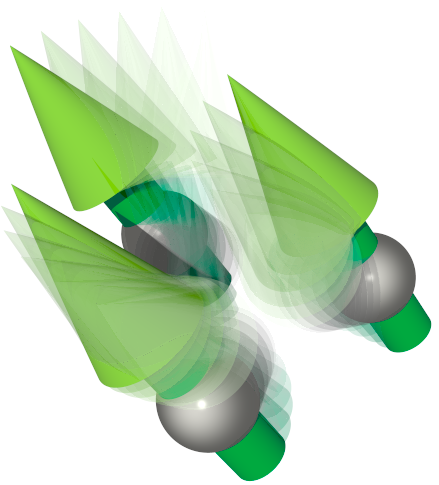}
\caption{(Color online) Conceptual figure of the breathing mode in a trimer of magnetic moments. The spins and atoms are represented as green arrows and gray balls, respectively.} 
\label{fig:principle} 
\end{figure}

As already pointed out, dissipation is one of the consequences of these couplings. Since the electron motion is several orders of magnitude faster than both spin and lattice motion, it can, for some purposes, be integrated out. \cite{Antropov1996,Skubic2008,Eriksson2017asd} The spin and lattice degrees of freedom, however, occur at a much slower and roughly equivalent time scale and need to be addressed in a unified way, self-consistently. \cite{Antropov1996,Omelyan2001,Ma2008} Figure~\ref{fig:principle} shows a schematic picture of coupled spin-lattice dynamics. It has been demonstrated that the exchange interactions between atoms in several magnetic materials can depend strongly on the local atomic environment, and vice-versa; the chemical interaction may depend on magnetic configuration.\cite{Shimizu1981,Sabiryanov1995,Sabiryanov1999,Bonetti2016} Hence, both the magnon and phonon spectrum and lifetimes in a material may depend on the configuration of magnetic state or the displacement of atoms.\cite{Hasegawa1987,Koermann2014}

Several studies point at the importance of phonon-magnon coupling in a number of dynamical processes such as  demagnetization processes,\cite{Carpene2008,Wietstruk2011,Reid2018} thermal conductivity,\cite{Boona2014,Sanders1977}
magneto-acoustics,\cite{Kim2012prl,Scherbakov2010,Henighan2016} 
and the spin-Seebeck effect.\cite{Jaworski2011,Man2017,Maehrlein2017pp} 
The interaction between spin and lattice motion is also central for phenomena observed in magnetoelectric and in multiferroic materials,\cite{Pimenov2006,ValdesAguilar2007,Sushkov2008,Takahashi2012,Kubacka2014,Wang2012,Tokura2014,Bossini2018} 
magnetocaloric materials,\cite{Tishin2014} 
skutterudites,\cite{Rasander2015} and
antiferromagnetic insulator materials for spintronic devices.\cite{Aytan2017} Among recent developments on methods for modelling of phonon-magnon coupling can be mentioned a novel combination of atomistic spin dynamics and ab initio molecular dynamics, applied to the paramagnetic phase of the magnetic insulator CrN,\cite{Stockem2018pp} and a scheme for massively parallel symplectic integration of spin-lattice dynamics equations of motion.\cite{Tranchida2018accepted}

In the present work, we describe a general method for the simulation of coupled spin-lattice dynamics, where all information needed for the effective spin-lattice dynamics Hamiltonian can be obtained from first-principles theory. We demonstrate the accuracy of the method by applying it to bcc Fe, as well as a selection of smaller clusters. The developed method is based on an atomistic approach and draws its strengths from the atomistic spin dynamics framework. In philosophy the method proposed here is similar to the early formulation of Ref. \onlinecite{Antropov1996}, but the practical details are naturally different. The interactions are described using a general Hamiltonian, with parameters computed using density functional theory. This hopefully provides a tool for analysis and even prediction of complex collective modes of magnetic materials that is a complement to experimental activities addressing these questions, for instance inelastic neutron scattering (INS) \cite{Squires2012itt} and resonant inelastic X-ray scattering (RIXS).\cite{2013msr,Ament2011,Schmitt2016} We note that the instrumentation and capabilities of these spectroscopies undergo a rapid development, for instance in form of prismatic analysers for neutron spectrometers \cite{Birk2014} for use in the CAMEA instrument \cite{Groitl2016} at the Paul-Scherrer Institute and in the BIFROST instrument \footnote{BIFROST: https://europeanspallationsource.se/article/bifrost-prismatic-approach-neutron-spectroscopy} commissioned for the European Spallation Source (ESS), and furthermore that INS and RIXS are complementary techniques that enable characterization of excitations throughout large parts of the Brillouin zone.\cite{Toth2016}

Using an empirical potential approach,\cite{Dudarev2005,Derlet2007} spin-lattice dynamics simulations of bcc Fe have been published by several groups \cite{Ma2008,Ma2012,Perera2014,Perera2017}. To put our method in perspective and on firm quantitative ground, we therefore selected to specifically address the spin-lattice interaction in bcc Fe as a test case. 

The rest of the paper is organized as follows.
In Section~\ref{sec:methods} we describe the Hamiltonian for coupled spin-lattice dynamics and the associated coupled equations of motion, techniques for calculation of the adiabatic magnon and phonon spectra, and a scheme for numerical integration of the coupled equations of motion. Section~\ref{sec:results} begins with a discussion of the dynamics of magnetic dimers, trimers, and quadmers, and continues with the results for bcc Fe. Finally, we discuss the applicability our of method and give an outlook in Sec.~\ref{sec:conc}. 

\section{Model and methods}
\label{sec:methods}

This section is split into five parts. First (Section~\ref{sec:SLDham}), we discuss the underlying Hamiltonian, which includes couplings within the spin and lattice reservoirs, respectively, and interactions between the spins and the lattice. Then, the couplings between the spin and lattice degrees of freedom are discussed in more detail in Section~\ref{sec:exchstr}. We motivate the equations-of-motion and the corresponding observables in Sections~\ref{sec:EOM} and \ref{sec:observ}, respectively. Finally, in Section~\ref{sec:DFT}, we describe how the coupling constants are obtained from density functional theory calculations.

\subsection{The spin-lattice Hamiltonian}
\label{sec:SLDham}

We consider the parametrised Hamiltonian
\bgea
\label{eq:HamSLD}
\mathcal{H}_{\rm SLD}=\mathcal{H}_{\rm S}+\mathcal{H}_{\rm L}+\mathcal{H}_{\rm LS},
\enea
formulated in terms of atomic magnetic moments $\{\bfm_i\}$, ionic displacements $\{\bfu_k\}$ and velocities $\{\bfv_k\}$. The first term is a Hamiltonian describing purely magnetic interactions. The second term contains energies associated with pure lattice displacements, and the third term couples the spin and lattice degrees of freedom. Indices $i,j$ run over atoms $1,\ldots,N_{\rm mag}$ with a finite magnetic moment, whereas indices $k,l$ run over all the atoms in the simulation cell $1,\ldots,N_{\rm all}$, i.e. non-magnetic as well as magnetic ions. Note that for the examples considered in the present paper, all ions are magnetic. Furthermore, $\alpha,\beta\in\{x,y,z\}$ denote the Cartesian components in spin space, and $\mu,\nu\in\{x,y,z\}$ denote the Cartesian components in real space. 

In the following, we consider contributions up to a combined order of four. The harmonic approximation to lattice dynamics is described by
\bgea
\label{eq:HamLL}
\mathcal{H}_{\rm LL}  = \frac{1}{2}\sum_{kl}\Phi_{kl}^{\mu\nu} u_k^{\mu} u_l^{\nu} 
+ \frac{1}{2} \sum_k M_k (v_k^{\mu})^2,
\enea
where the force constant tensor $\Phi_{kl}^{\mu\nu}$ is a rank 2 tensor in real space and $M_k$ is the mass of atom $k$. Since $\Phi_{kl}$ depends also on the magnetic configuration one can extend the expression above to 
\bgea
\label{eq:HamLLb}
\mathcal{H}_{\rm LL} = \frac{1}{2}\sum_{kl}\left(\Phi_{kl}^{\mu\nu}
+ \frac  {\partial \Phi_{kl}^{\mu\nu}} {\partial m^{\alpha}_i} \delta m^{\alpha}_i \right. \\ 
+ \left. \frac {\partial^2 \Phi_{kl}^{\mu\nu}} {\partial m^{\alpha}_i \partial m^{\beta}_j} \delta m^{\alpha}_i  \delta m^{\beta}_j \right) u_k^{\mu} u_l^{\nu} 
+ && \frac{1}{2} \sum_k M_k (v_k^{\mu})^2. \nonumber 
\enea
For some materials, the force constants depend significantly on the spin configuration as well as the configuration of nuclei. As an example we note that the traditional explanation for Invar alloys relies on the coupling between the force constants and the spin configuration,\cite{Schilfgaarde1999} as well as materials where many-body terms of the description of nuclear motion are needed. The main purpose of this paper is to outline a general formalism of coupled spin-lattice dynamics and to give examples of how the coupling modifies the dynamical properties in a few selected cases. In the illustration of the method we have for simplicity neglected the second and third terms inside the parenthesis, on the right-hand side of Eq. \ref{eq:HamLLb}, and instead kept a coupling term that originates from the Taylor expansion of exchange parameters (see below). The importance of various contributions to the coupling is materials dependent, but one may note that higher order contributions in the Taylor expansion are expected to be smaller. 

The bilinear spin Hamiltonian $\mathcal{H}_{\rm SS}$ contains Heisenberg exchange, Dzyaloshinskii-Moriya interaction and symmetric, anisotropic interactions that in a compact form can be expressed as 
\bgea
\label{eq:HamMM}
\mathcal{H}_{\rm SS}  = -\frac{1}{2}\sum_{ij} \mathcal{J}_{ij}^{\alpha\beta}(\{u_k^{\mu}\}) m_i^{\alpha} m_j^{\beta}.
\enea
The exchange tensor $\mathcal{J}_{ij}^{\alpha\beta}$ is a rank 2 tensor in spin space, with elements that in general have a dependence on the atomic displacements $\{u_k^{\mu}\}$ as well as the magnetic configuration. For clarity we specify in Eqn.4, the explicit dependence of tensor $\mathcal{J}_{ij}^{\alpha\beta}$ on $\{u_k^{\mu}\}$. The contributions to the mixed spin-lattice Hamiltonian $\mathcal{H}_{\rm LS}$ can then be obtained by expanding the bilinear magnetic Hamiltonian $\mathcal{H}_{\rm SS}$ up to second order in displacement, i.e. 
\bgea
\label{eq:HamMMexp}
\mathcal{H}_{\rm SS}  
&=& -\frac{1}{2}\sum_{ij}  \mathcal{J}_{ij}^{\alpha\beta} m_i^{\alpha} m_j^{\beta} -  \frac{1}{2}\sum_{ijk} \frac{\partial \mathcal{J}_{ij}^{\alpha\beta}}{\partial u_k^{\mu}} u_k^{\mu} m_i^{\alpha} m_j^{\beta} \nonumber\\
&& - \frac{1}{4}\sum_{ijkl} \frac{\partial^2  \mathcal{J}_{ij}^{\alpha\beta}}{\partial u_k^{\mu}\partial u_l^{\nu}} u_k^{\mu} u_l^{\nu} m_i^{\alpha} m_j^{\beta}.
\enea
Note that both in Eq. \ref{eq:HamMMexp} and Eq. \ref{eq:HamLLb} a term enters, that contains bilinear couplings in both spin- and lattice displacement. In one case it appears due to Taylor expansion of the Heisenberg exchange parameter in lattice displacement, and in the other it appears due to a Taylor expansion of the force constant in magnetic moments. In general both contributions are additive, as they have similar mathematical form. The relative importance of these two contributions is materials dependent and should preferably be calculated for the material one wants to investigate. The four-body interaction accounts for a renormalization of the phonon dispersion due to spin configuration. It also results in a renormalization of the magnon dispersion due to atomic displacements, and enables photon absorption by phonon-assisted multimagnon excitation.\cite{Lorenzana1995,Lorenzana1995a,Hellsvik2016}

Introducing the coupling constant
$\Gamma_{ijk}^{\alpha\beta\mu}=\nicefrac{\partial \mathcal{J}_{ij}^{\alpha\beta}}{\partial u_k^{\mu}}$ 
we may write the three-body interaction as
\bgea
\label{eq:HamMML}
\mathcal{H}_{\rm SSL}  &=& -\frac{1}{2}\sum_{ijk}\Gamma_{ijk}^{\alpha\beta\mu} m_i^{\alpha} m_j^{\beta} u_k^{\mu}.
\enea
This represents a spin-lattice coupling which is bilinear in spin and linear in displacement, i.e. $\Gamma_{ijk}^{\alpha\beta\mu}$ is a rank 3 tensor given by the direct product of a rank 2 tensor in spin space and a rank 1 tensor in orbital space. 

Defining 
$\Lambda_{ijkl}^{\alpha\beta\mu\nu} = \nicefrac{\partial^2 \mathcal{J}_{ij}^{\alpha\beta}}{\partial u_k^{\mu}\partial u_l^{\nu}}$ 
the four-body interaction reads 
\bgea
\label{eq:HamMMLL}
\mathcal{H}_{\rm SSLL}  &=& -\frac{1}{4}\sum_{ijkl}\Lambda_{ijkl}^{\alpha\beta\mu\nu} m_i^{\alpha} m_j^{\beta} u_k^{\mu} u_l^{\nu},
\enea
where the factor $1/4$ is due to that this interaction is bilinear both in spin and in displacements.
%
%
Taken together, the combined spin-lattice Hamiltonian reads
\bgea
\label{eq:HamSLDexp}
\mathcal{H}_{\rm SLD}  
&=& -\frac{1}{2}\sum_{ij}  \mathcal{J}_{ij}^{\alpha\beta} m_i^{\alpha} m_j^{\beta} -  \frac{1}{2}\sum_{ijk} \Gamma_{ijk}^{\alpha\beta\mu} u_k^{\mu} m_i^{\alpha} m_j^{\beta} \nonumber\\
&& -\frac{1}{4}\sum_{ijkl} \Lambda_{ijkl}^{\alpha\beta\mu\nu} u_k^{\mu} u_l^{\nu} m_i^{\alpha} m_j^{\beta} + \frac{1}{2}\sum_{kl}\Phi_{kl}^{\mu\nu} u_k^{\mu} u_l^{\nu} \nonumber\\
&& + \frac{1}{2} \sum_k M_k v_k^{\mu}v_k^{\mu}.
\enea
We note that in order to be even more general, higher coupling such as biquadratic exchange,\cite{Hayden2010,Hellsvik2014,Hellsvik2016} four-ring exchange,\cite{Fedorova2015} as well as third and fourth order\cite{Dove1993itl} interatomic lattice potential can be added, which is relatively straight-forward to do, and that for a very accurate description of the magnetic exchange dependence on the magnetic configuration needs to be considered.\cite{Szilva2013} Electrostatic contributions to the interatomic force field is also a relevant generalization to consider, since they can be important in polar materials, especially for excitations to the zone center. Likewise, magnetostatic interactions are sometimes of relevance. 

\subsection{Exchange striction}
\label{sec:exchstr}

The third-order spin-lattice coupling is considered usually in insulating magnets where the spin texture simultaneously breaks time and spatial reversion. This occurs, for instance, when describing ferroelectric polarization and multiferroic phases,\cite{Tokura2014} and it drives the magnetoelectric response in the electromagnetic field driven dynamics in the GHz and THz regime.\cite{Pimenov2006,ValdesAguilar2007,Sushkov2008,Takahashi2012,Kubacka2014} 

The isotropic (with regard to spin space) part of the $\Gamma_{ijk}^{\alpha\beta\mu}$ tensor is the exchange striction where the Heisenberg coupling between magnetic moments at $i$ and $j$ is modulated by the displacement of ion $k$. The antisymmetric anisotropic (with regard to spin space) part of the tensor represents the Dzyaloshinskii-Moriya interaction. This coupling will not be further discussed in the present paper, although the implementation discussed here does in principle allow for it.

In this paper we focus on the non-relativistic correction term to Heisenberg exchange
\begin{eqnarray}
\label{eq:HamExchStr}
\mathcal{H}_{\rm SSL}=-\frac{1}{2}\sum_{ijk} \vec{A}_{ijk}\cdot \bfu_k(\bfm_i\cdot\bfm_j),
\end{eqnarray}
in Eq.~\eqref{eq:HamMML} and for the simulations presented in Sec.~\ref{sec:results}, where atom $k$ could either coincide with one of the atoms $i$ and $j$ or be a distinct atom. The components of the exchange striction vector $\vec{A}_{ijk}$ relates to $\Gamma_{ijk}^{\alpha\beta\mu}$ according to
$\vec{A}_{ijk}^\mu=\Gamma_{ijk}^{xx\mu}=\Gamma_{ijk}^{yy\mu}=\Gamma_{ijk}^{zz\mu}$. The exchange striction coupling parameters are governed by certain symmetry rules: From the symmetry of Heisenberg exchange interaction
\bgea
\label{eq:exchsym}
\vec{A}_{ijk}=\vec{A}_{jik},
\enea
and in order to obey Newton's third law (forces are defined in Sec. \ref{sec:EOM}), the sum rule
\bgea
\label{eq:sumrule}
\vec{A}_{iji} = -\sum_{k\neq i}\vec{A}_{ijk}
\enea
has to hold. In different model approximations for the Heisenberg exchange, for instance the Ruderman–Kittel–Kasuya–Yosida (RKKY) type interaction \cite{Kundu:2015dn} or the effective model used by Ma \textit{et al.},\cite{Ma2017} the exchange interaction depends only on the distance $\vec{r}_{ij}$ between site $i$ and $j$, $\mathcal{J}_{ij}(\bfrr)=\mathcal{J}_{ij}(\bfrr_j+\bfu_j-\bfrr_i-\bfu_i)$, and, consequently, $\vec{A}_{iji}=|\vec{A}_{iji}| \bfr_{ij}$.
It obviously reflects the symmetry rule 
$\vec{A}_{iji}=-\vec{A}_{ijj}$, 
which is also valid in general for 
$\vec{A}_{ijk}$. 
It guarantees that the exchange striction force on atom $i$ and $j$ will cancel each other. The direction of the force caused by the exchange striction coupling is contained in 
$\vec{A}_{ijk}$ and is constrained by the point group symmetry of the crystal.

\subsection{The SLD Equations of motion}
\label{sec:EOM}

The coupled equations of motion for the spin-lattice system reads \cite{Ma2008}
\bgea
\frac{d\bfm_i}{dt} &=& -\frac{\gamma}{(1+{\alpha}^2)}\bfm_i \times \left(\bfbb_{i}+\bfbf_{i}\right) \label{eq:SLDeom1} \\
&&-\frac{\gamma}{(1+{\alpha}^2)}\frac{\alpha}{m_i}\bfm_i\times\left(\bfm_i\times[\bfbb_{i}+\bfbf_{i}]\right),\nonumber\\
\frac{d\bfu_k}{dt} &=& \bfv_k, \label{eq:SLDeom2}\\
\frac{d\bfv_k}{dt} &=& \frac{\bfff_k}{M_k} + \frac{\bffff_k}{M_k} - \nu\bfv_k, \label{eq:SLDeom3}
\enea
when expressed in the form of Langevin equations. Here, the effective magnetic field is obtained from the SLD Hamiltonian in Eq. \eqref{eq:HamSLD} as $\bfbb_i=-\partial{\mathcal{H}_{\rm SLD}}/{\partial \bfm_i}$ and the effective interatomic force field is determined from $\bfff_k=-\partial{\mathcal{H}_{\rm SLD}}/{\partial\bfu_k}$. $M_k$ is the mass of the atom at site $k$, $m$ is the saturation magnetization, $\gamma$ is the gyromagnetic ratio while $\alpha$ and $\nu$ are scalar (isotropic) damping constants. The stochastic fields $\bfbf_i$ and $\bffff_k$ are assumed to obey white noise properties, which implies that $\langle\bfbf_i(t)\bfbf_j(t^\prime)\rangle=2 D_M\delta_{ij}\delta(t-t^\prime)$ and $\bffff_k(t)\bffff_l(t^\prime)=2 D_L\delta_{kl}\delta(t-t^\prime)$. From the fluctuation-dissipation theorem it follows that $D_M = \alpha k_B T / \gamma m$ 
and $D_L=\nu M k_B T$.\cite{Ma2008}

The coupled partial differential equations are numerically solved using the method by Mentink \textit{et al.}~\cite{Mentink2010} combined with the Gr{\o}nbech-Jensen-Farago Verlet-type method,\cite{Gronbech-Jensen2013} or with a fixed-point scheme for implicit midpoint method. The methods require numerical step width of the order $\unit[10^{-15}]{s}$ down to $\unit[10^{-16}]{s}$. Details about the algorithms are provided in Appendix \ref{append:numint}.

Finally, we emphasize the fundamental difference between the here presented method and the one proposed by Ma \textit{et al.} \cite{Ma2008,Ma2012} and Perera \textit{et al.} \cite{Perera2017}: \textit{i)} All sets of parameters $\{J\}$,$\{\phi\}$, and $\{\Gamma_{ijk}^{\alpha\beta\mu}\}$ 
are determined from first principles and are not obtained from an effective potential or exchange model; \textit{ii)} we consider an established parametrisation of the lattice potential as presented in Eq.~\eqref{eq:HamLL}, which is directly available from standard first-principles tools (see Section~\ref{sec:DFT}); \text{iii)} the exchange striction term contains also couplings for $k\neq i,j$ and, consequently, will be highly applicable for magnets showing super- and double exchange mechanism for the magnetic coupling.

\subsection{Observables}
\label{sec:observ}

The primary output of a spin-lattice dynamics simulation are trajectories in time of the system variables $\{\bfm_i\}$, $\{\bfu_k\}$ and $\{\bfv_k\}$. In order to sample spatial and temporal fluctuations of the spins and the ions, we define the space- and time-displaced pair correlation functions
\bgea
C_{(\bfm)}^{\alpha \beta}(\mathbf{r},t-t_0)
&=& \frac{1}{N}\sum_{\substack{i,j ~ \text{where} \\\ ~\mathbf{r}_i-\mathbf{r}_j=\mathbf{r}}} \langle m_i^\alpha(t) m_j^\beta(t_0) \rangle, \label{eq:cmrt}\\
C_{(\bfu)}^{\mu \nu}(\mathbf{r},t-t_0) &=&\frac{1}{N}\sum_{\substack{i,j ~ \text{where} \\\ ~\mathbf{r}_i-\mathbf{r}_j=\mathbf{r}}} \langle u_i^\mu(t) u_j^\nu(t_0) \rangle, \label{eq:curt}\\
C_{(\bfv)}^{\mu \nu}(\mathbf{r},t-t_0) &=&\frac{1}{N}\sum_{\substack{i,j ~ \text{where} \\\  ~\mathbf{r}_i-\mathbf{r}_j=\mathbf{r}}} \langle v_i^\mu(t) v_j^\nu(t_0) \rangle .\label{eq:cvrt}
\enea
Equations~\ref{eq:cmrt}-\ref{eq:cvrt} can thus describe how the magnetic-, displacement- and velocity correlations evolves both in space and over time. It would of course be natural to investigate coupled modes, via coupled correlations, e.g., between moment and displacement in a similar way as outlined in Eqs.~\ref{eq:cmrt}-\ref{eq:cvrt}.
In the context of the simulations of bcc Fe in the present paper, the more relevant property is however obtained by a Fourier transform over space and time to give the dynamic structure factor for spin, displacement, and velocities. Defining relative time $\tau=t-t_0$ we obtain
\bgea
S_{(\bfm)}^{\alpha \beta}(\mathbf{Q},\omega)=\frac{1}{N\sqrt{2\pi}}\sum_{\mathbf{r}} e^{i\mathbf{Q}\cdot\mathbf{r}} \int_{-\infty}^\infty C_{(\bfm)}^{\alpha \beta}(\mathbf{r},\tau) \,d\tau,  \label{eq:msqw} \\
S_{(\bfu)}^{\alpha \beta}(\mathbf{Q},\omega)=\frac{1}{N\sqrt{2\pi}}\sum_{\mathbf{r}} e^{i\mathbf{Q}\cdot\mathbf{r}} \int_{-\infty}^\infty C_{(\bfu)}^{\alpha \beta}(\mathbf{r},\tau) \,d\tau, \label{eq:usqw} \\
S_{(\bfv)}^{\alpha \beta}(\mathbf{Q},\omega)=\frac{1}{N\sqrt{2\pi}}\sum_{\mathbf{r}} e^{i\mathbf{Q}\cdot\mathbf{r}} \int_{-\infty}^\infty C_{(\bfv)}^{\alpha \beta}(\mathbf{r},\tau) \,d\tau, \label{eq:vsqw}
\enea
which are closely related to what is measured by inelastic neutron or electron scattering experiments. The dynamic structure factors are naturally analyzed in terms of the differential cross section \cite{Squires2012itt} which for many materials is proportional to the dynamical structure function. This means that by simulating the dynamical structure factor, the relation between momentum transfer $\mathbf{Q}$ and frequency $\omega$ for magnons and phonons in the material can be obtained.

For systems lacking periodicity, such as the small dimer, trimer, and quadmer clusters that are also considered in this work, the $\mathbf{Q}$-dependance of the structure factor is in principle undefined. Instead, the spatially dependent excitation spectra of these finite systems can be obtained by performing the Fourier transform over time only. Summing the resulting correlation function over all sites results in the total excitation spectra of the system.

The spin dynamical structure factor accurately describe magnon dispersions, especially in thin films,\cite{Bergqvist2013} since it properly takes into account magnon-magnon scattering properties and damping at finite temperatures. On the other hand, in the limit of very low temperatures and damping, the magnon dispersion is more easily obtained through the adiabatic linear spin wave theory.\cite{Singer:2007hi}

Let us first focus on the spin-degree of freedom and the adiabatic magnon spectra for the collinear magnetic case with a system consisting of 1 atom/cell such as bcc Fe. Then the spatial Fourier transform of the exchange interactions reads 
\bgea
\mathcal{J}(\mathbf{Q})=\sum_{j\ne0}
\mathcal{J}_{0j} e^{i\bfqq\cdot(\bfrr_j-\bfrr_0)},
\enea
where $\mathcal{J}_{0j}$ is the exchange interaction between magnetic atoms at site $0$ and $j$, respectively. Note that here the index $j$ runs over all magnetic sites with the origin at $\bfrr_0$. The spin wave energies $\omega(\mathbf{Q})$ will then be given by the following expression \cite{Halilov1998} 
\bgea
\label{eq:AMS}
\omega(\mathbf{Q})=\frac{4}{M}\left( \mathcal{J}(\mathbf{0})-\mathcal{J}(\mathbf{Q}) \right). 
\enea
Generalization can be done towards multi-sublattice systems, see e.g. Ref.~\onlinecite{Eriksson2017asd}, and using Bogoliubov transformation \cite{Colpa1978} towards general non-collinear formulation.\cite{Toth2015,Yadav2017pp}

For lattice degree-of-freedom, the reciprocal space dynamical matrix $D_{s\beta t\nu}(\bfqq)$ is related to the force constant matrix in real space by the mass normalised Fourier transform
\bgea
\label{eq:dynmat}
D^{\mu\nu}_{st}(\bfqq)=\frac{1}{\sqrt{M_sM_t}}\sum_l\Phi^{\mu\nu}_{0slt}e^{[i\bfqq\cdot(\bfrr_l-\bfrr_0)]}
\enea
where $M_s$ is the mass of atom $s$ in the unit cell. Given the translational symmetry of the crystal, it is enough to sum only over $l$ in all the $N_P$ primitive cells in the supercell. For the $\Gamma$ point, the Fourier transform is a plain sum over all repetitions of the primitive cell, both the ones contained in the Wigner-Seitz cell of the simulation supercell and, due to the periodic boundary conditions, the ones outside it.\cite{Wang2010a} Note that in linear response density functional perturbation theory computation of phonons, it is actually the elements of the dynamical matrix that are calculated from which the interatomic force constants can be calculated by the inverse Fourier transform.\cite{Baroni2001} Solving the eigenvalue problem
\bgea
\label{eq:phondisp}
{\rm det}[D_{s\mu t\nu}(\bfqq)-\omega^2(\bfqq)]=0
\enea
the $3N$ phonon modes (eigenvectors) and frequencies (the square roots of the occasionally degenerate eigenvalues $\omega^2$) are obtained for a given $\bfqq$ vector.
Opposite to the dynamic structure factor (Eqs. \ref{eq:msqw}-\ref{eq:vsqw}), the adiabatic spectra described above do not directly account for the coupling between the spin- and lattice reservoir. A possible way forward in this regard is to replace the exchange interaction and force constant matrix in Eqns.21 and 23, with corresponding Taylor expanded entities defined in Eqns.5 and 3, respectively. This can also be handled as described e.g in Refs. \onlinecite{White:1965ji,Kim2007,Ruckriegel2014,Toth2016}, which is, however, beyond the 
scope of the present paper. 

\subsection{DFT calculations}\label{sec:DFT}
Aiming for a first-principles description of the coupled spin-lattice dynamics in bcc Fe, we calculated the coupling constants that occur in Eq.~\ref{eq:HamSLDexp} by means of density functional theory (DFT) calculations. 

The harmonic force constants $\Phi_{kl}^{\mu\nu}$ were calculated with the finite displacement method using the Vienna {\it ab initio} Vienna simulation package (VASP)\cite{Kresse1996a,Kresse1996} and the Phonopy \cite{Togo2015} software. The VASP calculations were performed using the projector augmented wave method\cite{Bloechl1994,Kresse1999} and the local density approximation as exchange-correlation functional. A $6\times6\times6$ supercell was used for bcc Fe. We employed a plane-wave energy cutoff of 600 eV and a $\Gamma$-centered $\textbf{\emph{k}}$-points mesh of size $4\times4\times4$.

In order to achieve a complete first-principles spin-lattice model, we also approach the Heisenberg interaction and the exchange striction from DFT. To this end, we applied the full-potential linear muffin-tin orbitals (FP-LMTO) method as implemented in the RSPt software~\cite{Wills2010fes}. The maximum value of the angular momentum used for the angular ($l$) decomposition of the charge density and the potential inside the muffin-tin spheres was taken equal to $l_{max}$ = 12. Three kinetic energy tails were used for the description of the states in the interstitial region: -0.3, -2.3, and -1.5 Ryd. Within this setup, we calculated the Heisenberg exchange coupling $\mathcal{J}_{ij}$ via the Liechtenstein-Katsnelson-Antropov-Gubanov (LKAG) formalism \cite{Liechtenstein1984,Liechtenstein1987}. For the actual implementation into the RSPt code we refer the reader to Ref. \onlinecite{Kvashnin2015}. Recently, this method has been successfully applied to strongly correlated systems such as NiO\cite{{Kvashnin2015}} and BiFeO$_3$~\cite{Paul2018}.

The magnetic exchange interactions $\mathcal{J}_{ij}$ have been calculated with RSPt for a primitive cell and for a $2\times2\times2$ supercell for which identical $\mathcal{J}_{ij}$ came out. In order to calculate first order exchange striction, $\mathcal{J}_{ij}$ have been calculated 
for the same $2\times2\times2$ cell but with one atom displaced with a finite displacement $\Delta$ along $\vec{e}_{\Delta}$. A $\textbf{\emph{k}}$-point mesh of $30 \times 30 \times 30$ was employed for the supercell calculations. The so obtained set of exchange couplings have a lower symmetry. For instance, the set of eight equivalent nearest neighbor couplings in the bcc structure are broken up to sets of 1, 3, 3, 1 degenerate couplings for a distortion along the [100] axis (in bcc lattice vectors, i.e. along the [111] for Cartesian axis), and sets of 4, 4 couplings for a distortion along the $[110]$ axis (Cartesian $[100]$). For the bcc structure we have compared carefully that for different distortion directions, the symmetry lowering for the $\mathcal{J}_{ij}$ up to the fourth coordination shell is identical to the symmetry lowering of the crystal itself.

Since anisotropic exchange parts are not considered, the tensor $\mat{\Gamma}_{ijk}$ becomes a vector $\vec{A}_{ijk}$. Similar to the force constants and magnetic exchange, $\vec{A}_{ijk}$ fulfill point group symmetries, in  particular for bcc Fe the 48 symmetry operations of space group number 229 (Im$\bar{3}$m). Furthermore, the exchange striction energy $E_{ijk}$ related to the sites $i$, $j$, and $k$ is isotropic. Consequently, we obtain $\vec{A}_{ijk}=\vec{A}_{jik}$ from $E_{ijk}=E_{jik}$, which is furthermore caused by the isotropic properties of the magnetic exchange. In our spin-lattice dynamics simulations we even have to guarantee that the centre of mass is not drifting, which is guaranteed when also the spin-lattice couplings fulfill Newton's third law. 

Using a finite difference method involving the non-displaced set $\{\mathcal{J}^0\}$ and a displaced set $\{\mathcal{J}^{\vec{e}_{\Delta}}\}$, we obtain the directional derivative $\Gamma^{\Delta}_{ijk}$. The gradient $\nabla_k\mat{\mathcal{J}}_{ij}$ is finally constructed from the directional derivative definition and out of three different sets $\{\mathcal{J}^{\vec{e}^{\nu}_{\Delta}}\}$ of independent directions $\vec{e}^{\nu}_{\Delta}$, where $\nu=1,2,3$, but the same displacement strength $\Delta$. We have chosen $\Delta$ to be 0.003$a_0$, 0.002$a_0$,0.001$a_0$, and 0.00001$a_0$, where $a_0$ is the Bohr radius. In order to fulfill the finite displacement criteria, $\Delta$ is interpolated to zero by Hermite interpolation. Numerical noise is reduced by applying various symmetries, as discussed above. We apply iteratively the above mentioned symmetry operations until we reach convergence. It is important to mention that in the last step Newton's third law has to be enforced.

\section{Results}\label{sec:results}

This section is divided into two parts. In Section~\ref{sec:exch234} we discuss the application of our method to low-dimensional model systems and discuss symmetry-related issues of the three-body exchange coupling. 
(Note that we have selected, for simplicity, to neglect four-body interaction in the present work.) Section~\ref{sec:bccFe} deals with applications to real materials, in particular to bcc Fe. All required parameters are calculated from first principles. We present quasi-particle dispersion relations and discuss the role of the three-body interaction in the spectra at various temperatures.

\subsection{Exchange striction in 2-, 3- and 4-site systems}
\label{sec:exch234}

As conceptual examples for our method, we perform coupled spin-lattice dynamics simulations for systems consisting of two (dimer), three (trimer), and four (quadmer) atoms. If not mentioned, we neglect energy dissipation in our model and, consequently, the total energy has to be conserved. Furthermore, we account only for the isotropic part of the magnetic exchange tensor, namely the Heisenberg exchange, but note that anisotropy in general is of significant importance in low-dimensional systems \cite{Bergqvist2013,Bergman2016,Smogunov2007,LeBacq2002,Delin2004,Menendez2000}.

\subsubsection{Dimer}
\label{sec:exchstrdimer}

\begin{figure}
\includegraphics[width=0.55\textwidth]{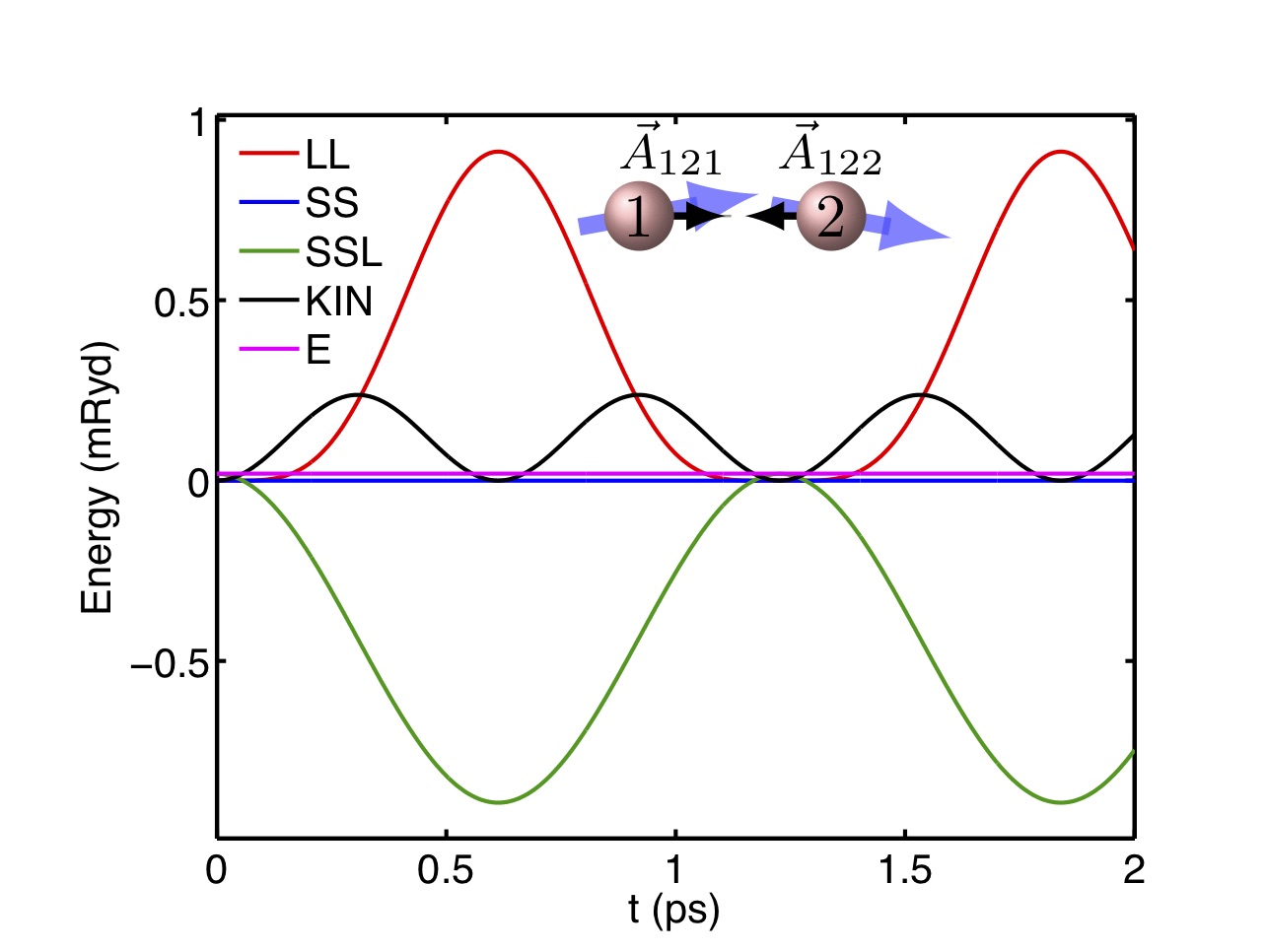}
\caption{(Color online)  Energy trajectories at $T=0$ K of a dimer oriented along the $x$-direction with Hamiltonian according to Eqs. \ref{eq:HamDimer}-\ref{eq:HamDimer3}. The inset shows a conceptual figure of the exchange striction coupling constants $A_{12k}$, $k=1,2$ (black arrows) in a dimer. The atoms are represented by golden balls, where the initial magnetic moment configuration in the dimer is given by purple arrows.}
\label{fig:sldenergy1}
\end{figure}

In this model system, we consider a dimer where the two sites are denoted by 1 and 2 (see the inset of Fig.~\ref{fig:sldenergy1}) and we have set $M_1=M_2=$ atomic mass units, and $m_1=m_2=1$ $\mu_B$. The simplicity of this system allows to provide explicit expressions for the Hamiltonian, the effective magnetic fields and the interatomic forces. The four parts of the dimer Hamiltonian 
\begin{eqnarray}\label{eq:HamDimer}
\mathcal{H}^{\mathrm{dimer}} &=& \mathcal{H}_{\mathrm{LL}}^{\mathrm{dimer}} + \mathcal{H}_{\mathrm{SS}}^{\mathrm{dimer}} +
\mathcal{H}_{\mathrm{SSL}}^{\mathrm{dimer}} +
\mathcal{H}_{\mathrm{KIN}}^{\mathrm{dimer}}
\end{eqnarray}
reads 
\begin{eqnarray}\label{eq:HamDimer2}
\mathcal{H}_{\mathrm{LL}}^{\mathrm{dimer}} &=& \frac{1}{2}\Phi_{11}^{\mu\nu} u_1^{\mu} u_1^{\nu} + \frac{1}{2}\Phi_{12}^{\mu\nu} u_1^{\mu} u_2^{\nu}, \nonumber\\
&& + \frac{1}{2}\Phi_{21}^{\mu\nu} u_2^{\mu} u_1^{\nu} + \frac{1}{2}\Phi_{22}^{\mu\nu} u_2^{\mu} u_2^{\nu}, \\
\mathcal{H}_{\mathrm{SS}}^{\mathrm{dimer}} &=& -\frac{1}{2}J_{12}\bfm_1\cdot\bfm_2 - \frac{1}{2}J_{21}\bfm_2\cdot\bfm_1, \\
\mathcal{H}_{\mathrm{SSL}}^{\mathrm{dimer}}
&=&  \frac{1}{2} A_{121}^{\mu}(\bfm_1\cdot\bfm_2)u_1^{\mu} + \frac{1}{2} A_{122}^{\mu}(\bfm_1\cdot\bfm_2)u_2^{\mu} \\
&& + \frac{1}{2} A_{211}^{\mu}(\bfm_2\cdot\bfm_1)u_1^{\mu} + \frac{1}{2} A_{212}^{\mu}(\bfm_2\cdot\bfm_1)u_2^{\mu}, \nonumber\\
\mathcal{H}_{\mathrm{KIN}}^{\mathrm{dimer}} &=& \frac{1}{2} M_1 v_1^{\mu}v_1^{\mu} +  \frac{1}{2} M_2 v_2^{\mu}v_2^{\mu}.\label{eq:HamDimer3}
\end{eqnarray}
In particular for the dimer, we choose the magnetic interaction to be $J=\unit[0.1]{mRyd}$ and the harmonic atomic force constants uniaxial with $\phi_{12}=-\unit[1]{Ryd \angstrom^{-2}}$. The three-body interaction is introduced along the bond and is set to $\unit[1]{mRyd \angstrom^{-1}}$. The scalar product of the two moments in the dimer is, when damping is ignored, a constant of motion and hence also the exchange energy will be constant (see Fig.~\ref{fig:sldenergy1} blue line). 

The evolution of the energy origins from the corresponding harmonic interatomic forces
\begin{eqnarray}
F_{\mathrm{LL},1}^{\mathrm{dimer},\mu} &=&  -\Phi_{11}^{\mu\nu} u_1^{\nu} -\Phi_{12}^{\mu\nu} u_2^{\nu},
\label{eq:FDimerLL2}
\end{eqnarray}
and interatomic forces from the three-body exchange 
\begin{align}
 F_{\mathrm{SLL},1}^{\mathrm{dimer},\mu} = -A_{121}^{\mu}(\bfm_1\cdot\bfm_2),
\label{eq:FDimerMML2}   
\end{align}
where we used the symmetry relation of the force constants $\Phi_{kl}^{\mu\nu}=\Phi_{lk}^{\nu\mu}$ and the exchange striction terms $A_{121}^{\mu}=A_{211}^{\mu}$.

\begin{figure}
\includegraphics[width=0.50\textwidth]{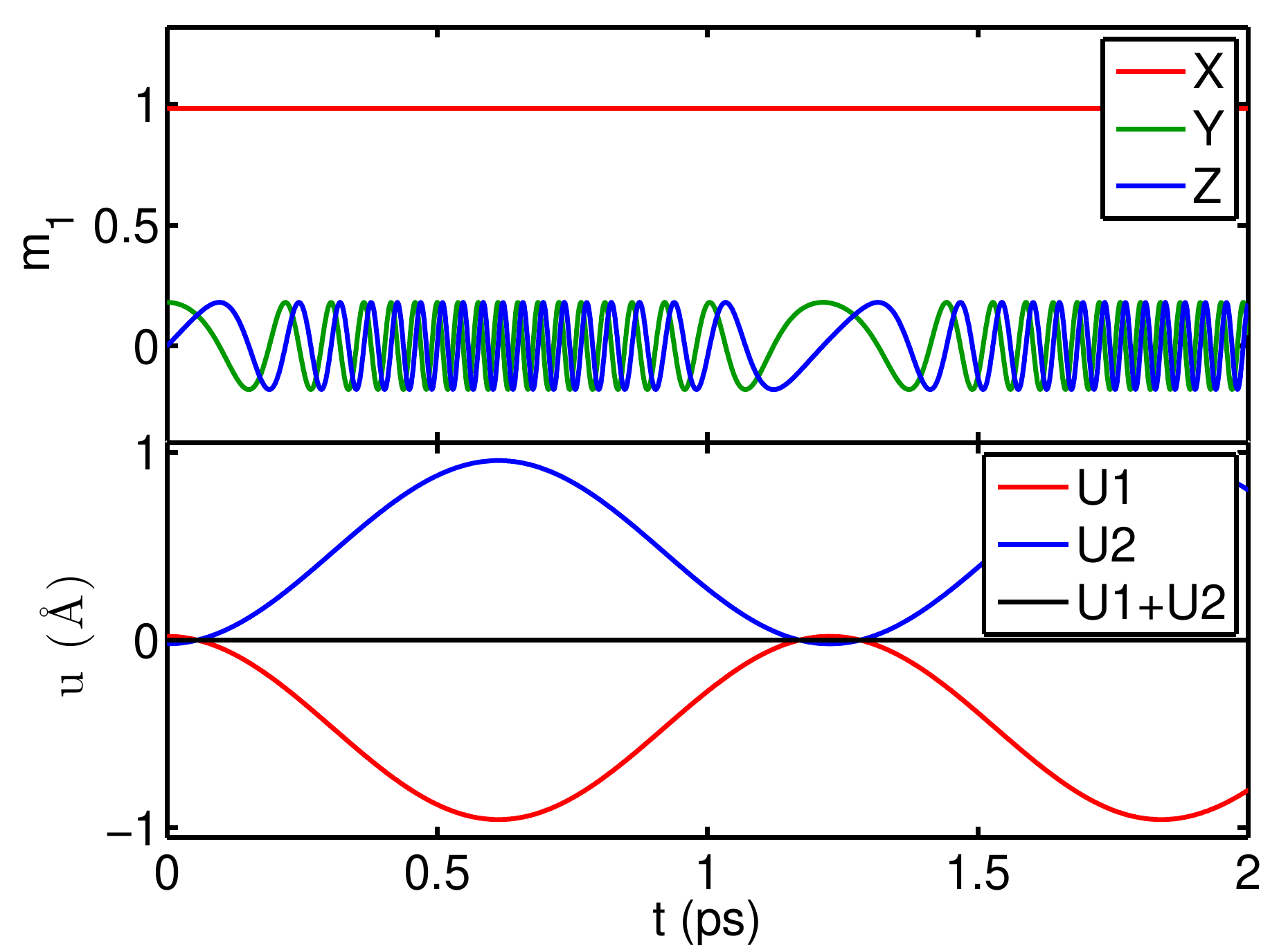}
\caption{(Color online) Trajectories at $T=0$ K of a dimer oriented along the $x$-direction. The slow ionic motion couples over the exchange striction to the spin system and induces a modulation of the frequencies of spin precession. (upper panel) The Cartesian components of $\bfm_1$, labeled in the figure with $X$, $Y$, and $Z$. (lower panel) The displacement of ion 1 and 2 along the dimer bond axis.}
\label{fig:dispmagtraj1}
\end{figure}

Likewise, the magnetic degree of freedom is driven by the magnetic exchange field
\begin{align}
\bfbb_{\mathrm{SS},1} = J_{12}\bfm_2,
\end{align}
and exchange striction field
\begin{align}
\bfbb_{\mathrm{SSL},1} = -A_{121}^{\mu}\bfm_2 u_1^{\mu} -A_{122}^{\mu}\bfm_2 u_2^{\mu}.
\label{eq:BDimerMML}   
\end{align}
Here, we applied the isotropy property of the magnetic exchange $J_{12}=J_{21}$. Equations~\eqref{eq:FDimerLL2}-\eqref{eq:BDimerMML} show that the direction of $\vec{F}_{\mathrm{SLL},1}^{\mathrm{dimer}}$ is only dictated by the coupling constant, where the amplitude is also related to the relative angle between the magnetic moments. Hence, in the absence of Gilbert damping that strives to align the spins  $\vec{F}_{\mathrm{SLL},1}^{\mathrm{dimer}}$ will be also a constant of motion in the case of a dimer. For the effective magnetic field, the three-body term only scales the field strength. The exchange striction term conserves the center of mass and, consequently, $\vec{u}_1=-\vec{u}_{2}$ and $\vec{A}_{121}=-\vec{A}_{122}$. Thus, the case $J_{12}<2\vec{A}_{121}\cdot\vec{u}_1$ is of high interest, since the effective exchange switches from a ferromagnetically to an antiferromagnetically coupled system. For the dimer, however, this will only change the direction of precession locally in time. 

The fields and forces in Eqs.~\eqref{eq:FDimerLL2}-\eqref{eq:BDimerMML} finally lead to the evolution of the spin and lattice degree-of-freedom as shown in Fig.~\ref{fig:sldenergy1}, for the different contributions to the energy, and in Fig.\ref{fig:dispmagtraj1} for the displacement and moment direction. The simulations reproduce the conservation of the center of mass (Fig.~\ref{fig:dispmagtraj1} lower panel) as well as the relative angle between the magnetic moments. The initial conditions for the displacement in the dimer are set to be $\unit[0.02]{\angstrom}$ ($2\%$ of the lattice constant). It should be noted that the oscillation is not around the equilibrium position ($\vec{u}=0$), but around a shifted position along the bond axis ($x$-axis).  On the other hand, the magnetic moments move only in the $yz$-plane as a result of the initial conditions (see Fig. \ref{fig:sldenergy1}, inset, purple arrows). The precession of the magnetic moment is seen to vary in time and to be largest when the displacement $\vec{u}_1$ is significant\footnote{See the supplementary movie}. This is a natural consequence of the exchange striction field, defined in Eqn.33.

\begin{figure}
\includegraphics[width=0.50\textwidth]{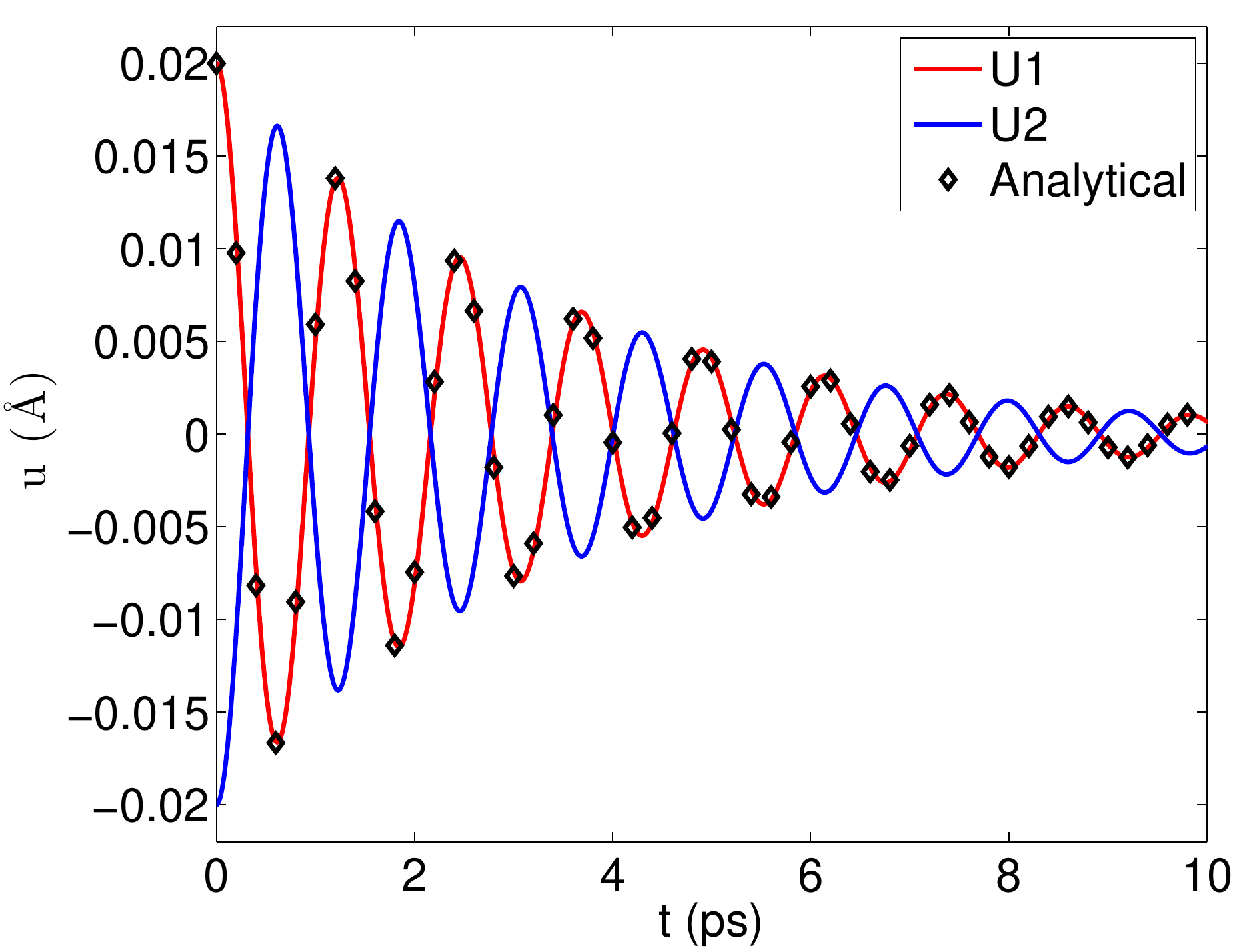}
\caption{(Color online) 
 Trajectories at $T=0$ K of the damped lattice motion of a dimer oriented along the $x$-direction. The analytical solution for atom 1 (black symbols) lies on top of the simulated trajectory (red line).}
\label{fig:xdisp}
\end{figure}

To further analyze the reliability of the method we also simulate the dynamics in the presence of a viscous damping $\nu=10^{-14}$ kg$/$s. The analytical solution of the damped 1D-harmonic oscillator (see e.g. Ref. \cite{Meriam1993em}) is compared with the numerical one and we obtain perfect agreement as shown in  Fig.~\ref{fig:xdisp}. The ions oscillate around the center of mass and the envelope of the trajectories decays exponentially in time. 

\subsubsection{Trimer}
\label{sec:exchstrtrimer}

\begin{figure}[ht]
\centering
\includegraphics[width=0.50\columnwidth]{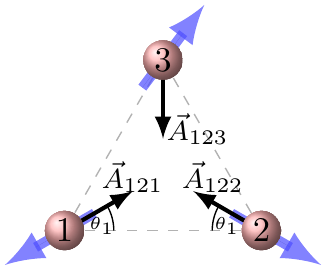}
\caption{(Color online) Conceptual figure of the exchange striction coupling constants $A_{12k}$, $k=1,2,3$ (black arrows) in a trimer. By symmetry, we have $A_{123}=-A_{121}-A_{122}$, and one free angle $\theta_1$ as a parameter.
The atoms are represented by red balls, where  the initial magnetic moment configuration in the trimer is given by blue arrows.}
\label{fig:trimer} 
\end{figure}

The three sites of a trimer are mutually nearest neighbors, which enables for a total of 18 possible $\vec{A}_{ijk}$ couplings. Respecting that symmetry under exchange of spin sites (Eq. \ref{eq:exchsym}), point group symmetry $D_{3h}$, and the sum rule (Eq. \ref{eq:sumrule}) should always hold, we consider the following case for the exchange striction (Fig.~\ref{fig:trimer}): Exchange striction vectors tilt away from the bond by angle $\theta_1$ (see Fig.~\ref{fig:trimer}): $\vec{A}_{iji}\nparallel\bfr_{ij}$ and, consequently, $\vec{A}_{ijk(k\neq i,j)}\neq\vec{0}$. Furthermore, $\vec{A}_{ij1} = -\sum_{l\neq 1}\vec{A}_{ijl}= -\sum_{l\neq 1}\vec{A}_{jil} = \vec{A}_{ji1}$. Note, that only for $\theta=\nicefrac{\pi}{6}$ we have $|\vec{A}_{iji}|=|\vec{A}_{ijk(k\neq i,j)}|$.

Note that we here refer to the amplitude of $\vec{A}_{iji}$ as $A$. Just as for the dimer case, we assume nearest neighbor coupling of $J=\unit[1]{mRyd}$. The irreducible part of the atomic force constants $\phi$ for $D_{3h}$ symmetry are in the notation of Refs.~\onlinecite{Cserti2004,Toth2016}, i.e.  $\Phi_2^{xx}=\unit[-0.25]{Ryd \angstrom^{-2}}$, $\Phi_2^{xy}=\unit[-0.43]{Ryd \angstrom^{-2}}$, and $\Phi_2^{yy}=\unit[-0.25]{Ryd \angstrom^{-2}}$. The mass of each atom is put to 1 atomic mass unit. 

\begin{figure}[ht]
\centering
\includegraphics[width=0.48\textwidth]{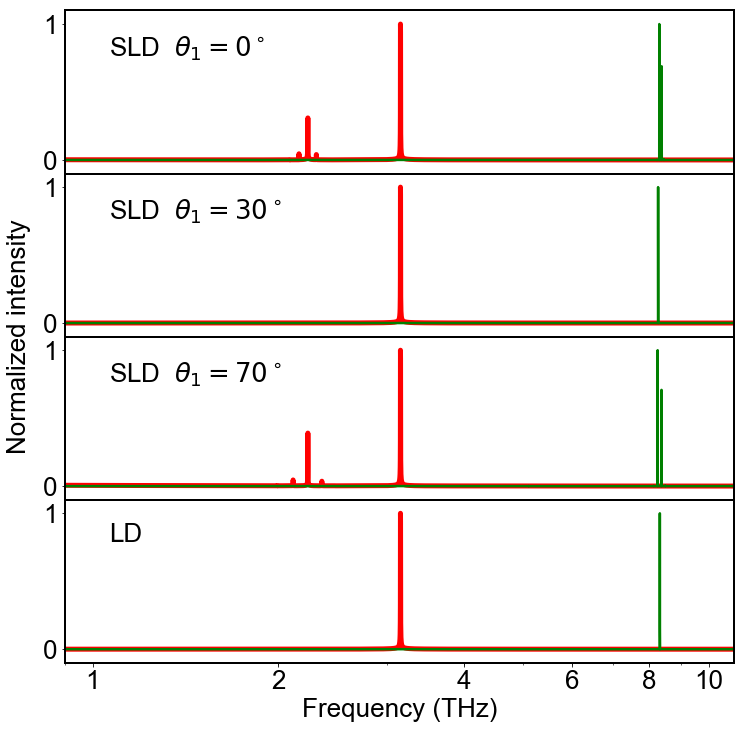}
\caption{(Color online) Excitation spectra for lattice displacements (red/thick lines) and magnetic oscillations (green/thin lines), for the trimer configurations. The top three panels corresponds to SLD simulations with $\theta_1=0^\circ$,  $\theta_1=30^\circ$ and $\theta_1=70^\circ$, as described in the text. The bottom panel corresponds to decoupled lattice-dynamics and spin-dynamic simulations. 
}
\label{fig:trimer_powdisp}
\end{figure}

In order to see direct effects of the $A$ coupling, we plot in Fig.~\ref{fig:trimer_powdisp}, the measured excitation spectra given by Eqn.18 and 19, obtained from simulations for three choices of $\theta_1$ (defined in Fig.5) with $A=\unit[0.1] {mRyd}$. For comparison, the excitation spectra for a reference system with $A=0$, i.e. without any coupling between the spin and lattice systems is also included in the figure. For the decoupled trimer, two significant peaks are present. The peak that is lower in energy, at $\unit[3.2] {THz}$, represents the lattice vibrations while magnetic fluctuations cause the other peak at  $\unit[8.2]{THz}$. In the case of a finite spin-lattice coupling we find that for two of the considered angles ($\theta_1=0^\circ$ and $\theta_1=70^\circ$), a very fine splitting of the magnetic energy level is noticable while a low-energy peak at $\unit[2.2] {THz}$ occurs for the lattice vibrations. Interestingly, for the third choice of angle, $\theta_1=30^\circ$, the difference compared to the decoupled system is found to be minimal. Frequencies such as the ones shown in Fig.6 are available through Raman spectroscopy, and an experimental detection of a spin-lattice coupling, via the split peaks in Fig.6 would be interesting.

\begin{figure}[t]
\centering
\includegraphics[width=0.35\textwidth]{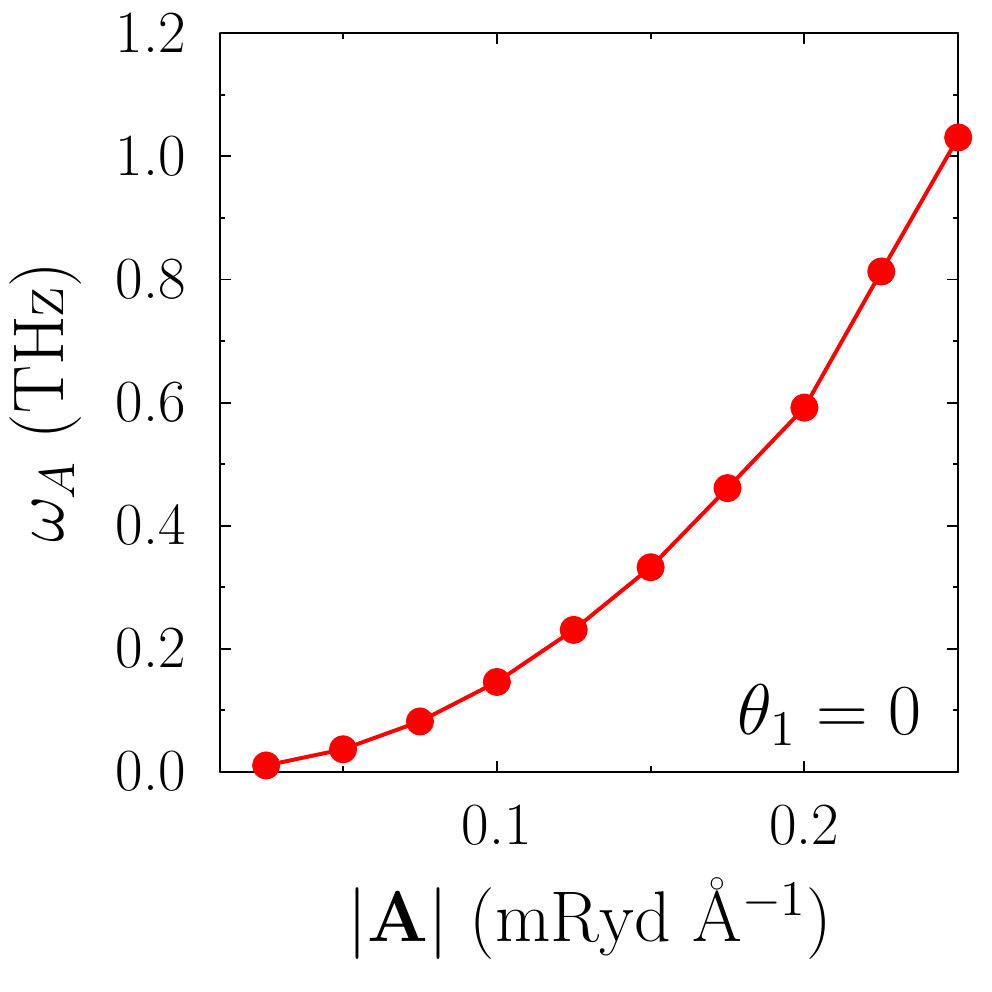}
\includegraphics[width=0.35\textwidth]{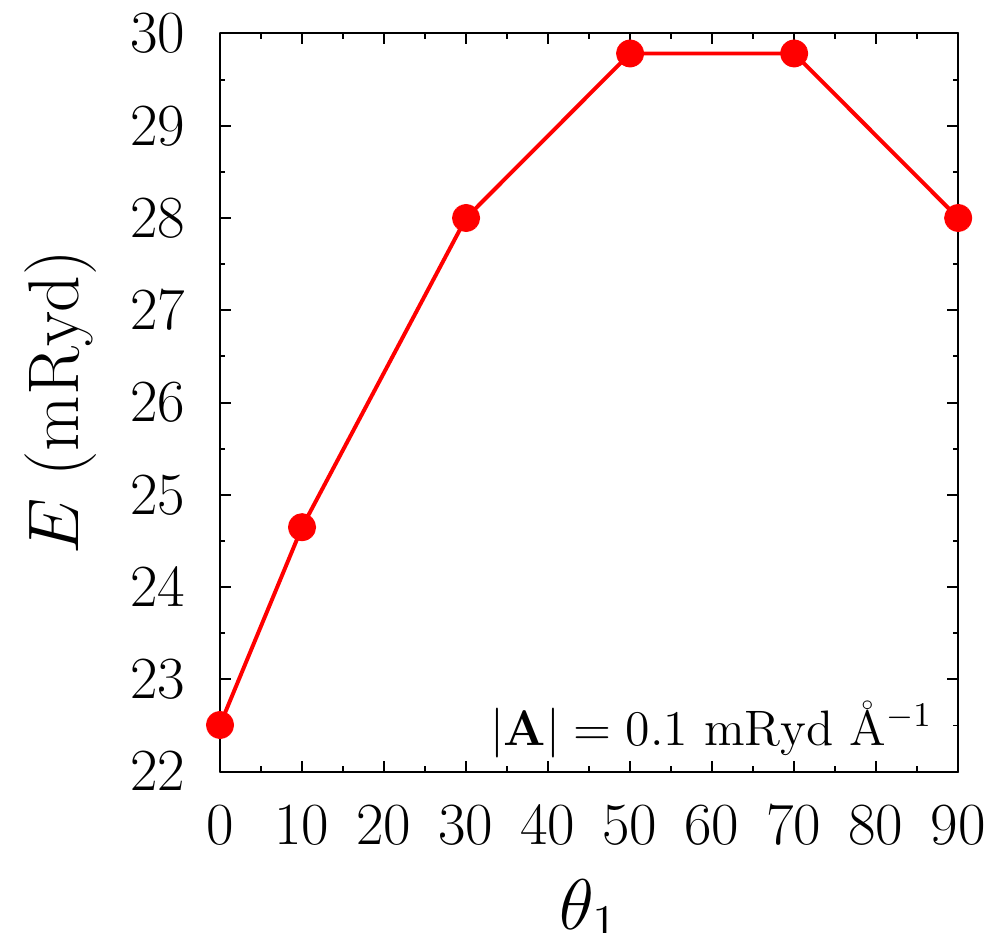}
\caption{(Color online) Spin-lattice dynamics of trimer. (upper panel) Envelope frequency of spin oscillations as function of $|A|$. (lower panel) Total energy as function of tilt angle of $A$.}
\label{fig:trimerprinciple}
\end{figure}

In our simulation we varied both the strength $A$ and the angle $\theta_1$ of the exchange striction coupling. The strength $A$ affects the frequency $\omega_A$ of an enveloping oscillation on top of the spin precession frequency $\omega_p$: $\omega_A$ scales quadratically with the strength $A$ (see Fig.~\ref{fig:trimerprinciple}(upper panel). Without energy dissipation, the magnetic energy is conserved, just as for the dimer case. The variations of the various energy contributions will be similar to the one in Fig.~\ref{fig:sldenergy1}. The total energy increases linearly with the strength $A$, but oscillates with $\theta_1$ which is related to the fixed initial spin configuration (see Figs.~\ref{fig:trimerprinciple}(lower panel)). 

\subsubsection{Quadmer}
\label{sec:exchstrquadmer}

\begin{figure}[ht]
\centering
\includegraphics[width=0.80\columnwidth]{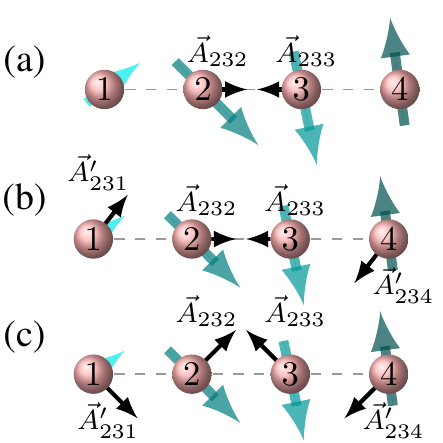}
\caption{\label{fig:quadmers}(Color online) Conceptual figure of the exchange striction coupling constants $A_{23k}$, $k=1,2,3,4$ (black arrows) in a quadmer. By symmetry, $A_{232}=-A_{233}$ and $A_{234}=-A_{231}$ are allowed. The atoms are represented by red balls. The initial, random magnetic moment configuration in the quadmer is given by blue arrows, where the length and orientation is related to the in-plane component, and the color to the $z$-component of the moment. The dotted line indicates the bond axis. 
}
\end{figure}

Although the trimer offers already rich phenomena, it addresses only nearest neighbor couplings, which are symmetry related and not independent. Contrary to the trimer, the sites of a 4-site system with periodic boundary condition (chain of atoms, see Fig.~\ref{fig:quadmers}(a), has both nearest neighbor (NN) and next nearest neighbors (NNN). Consequently, it is possible to have finite couplings $\vec{A}_{132}=\vec{A}_{312}=-\vec{A}_{134}=\vec{A}_{314}$ also when $\vec{A}_{iji}\parallel\bfr_{ij}$.

\begin{figure}[ht]
\centering
\includegraphics[width=0.48\textwidth]{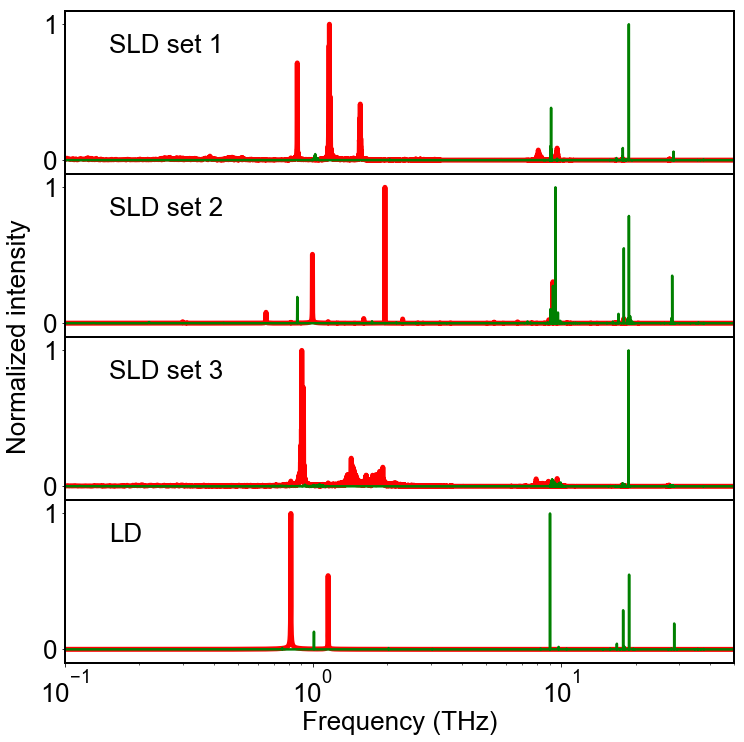}
\caption{(Color online) Excitation spectra for lattice displacements (red/thick lines) and magnetic oscillations (green/thin lines) for the considered quadmer clusters. The top three panels show the excitations for the SLD simulations corresponding to the three different cases discussed in the text. The bottom panel corresponds to decoupled LD and ASD excitations.
}
\label{fig:quadmer_powdisp}
\end{figure}

Here, we consider three different cases for the three-body interaction in the chain of four atoms.
\begin{enumerate}
    \item Exchange striction vectors parallel to bonds between spin $i$ and $j$ (Fig.~\ref{fig:quadmers}(a)): $\vec{A}_{iji}\parallel\bfr_{ij},\vec{A}_{ijk(k\neq i,j)}=\vec{0}$.
    \item Exchange striction vectors parallel to bonds between spin $i$ and $j$, but the second nearest neighbor coupling is different from zero (Fig.~\ref{fig:quadmers}(b)): $\vec{A}_{iji}\parallel\bfr_{ij},\vec{A}_{ijk(k\neq i,j)}\neq\vec{0}$. Thus, the sum over the second-nearest neighbor couplings has to cancel: $\sum_{k\in {\rm NNN}}\vec{A}_{ijk}=0$.
    \item Exchange striction vectors not parallel to bonds for all indices (Fig.~\ref{fig:quadmers}(c)): $\vec{A}_{iji}\nparallel\bfr_{ij},\vec{A}_{ijk(k\neq i,j)}\neq\vec{0}$
\end{enumerate}
For the Heisenberg exchange $J_{ij}$ and force constants $\phi_{ij}$ we include only nearest neighbor interactions of $J_{\rm NN}=\unit[1]{mRyd}$ and $\Phi_1^{xx}=\unit[-0.07]{Ryd \angstrom^{-2}}$. The atomic mass is put to 10 atomic mass units. In Fig.~\ref{fig:quadmer_powdisp} we show simulated excitation spectra for the tree sets of quadmer configurations discussed above, together with results from a decoupled simulation where all $\vec{A}_ijk=0$. Starting with the decoupled results in the lower panel of Fig.~\ref{fig:quadmer_powdisp} it is found that the lattice excitations are about one order of magnitude lower in energy compared to the magnetic excitations. The effect of finite spin-lattice couplings, shown in the top three panels in Fig.~\ref{fig:quadmer_powdisp}, is directly visible as lattice vibrations then contribute to high energy peaks, while magnetic fluctuations contribute to the lower energies. Compared to the pure lattice and magnetic excitation energies, the spectral weights of these coupled, or hybridized, excitations are however small. It can also be noticed that for the two sets of quadmer configurations with higher symmetry, (Sets 1 and 2), the excitation spectra has well defined peaks while for the lower symmetry configuration, set 3, certain peaks are more diffuse. This is most clearly visible for the lattice excitations at $\unit[2]{THz}$ and at $\unit[10]{THz}$ in the third panel of Fig.~\ref{fig:quadmer_powdisp}.

In effective pair-interaction models, as used for instance in Refs.~\onlinecite{Ma2012a,Perera2016}, only interactions between site $i$ and $j$ are possible. This situation corresponds to the system depicted in Fig.~\ref{fig:quadmers}(a) with its excitation spectra displayed in the top panel of Fig.~\ref{fig:quadmer_powdisp}. Allowing for three-body interactions, as in the system shown in Fig.~\ref{fig:quadmers}(b) results in the excitation spectra displayed in the second panel from the top in Fig.~\ref{fig:quadmer_powdisp}. Here, we have chosen that the NN and NNN three-body exchange have the same strength, e.g. $\unit[1]{mRyd\angstrom^{-1}}$. Comparing these two models we find for the spin excitations, that even though there is a difference in the distribution of spectral weight between the largest peaks, the actual spin excitation energies do not change much. There is however a larger difference noticeable for the lattice excitation energies, where both the position of the peaks and the spectral weights changes significantly when including NNN three-body exchange couplings. 

\subsection{Exchange striction in bcc Fe}
\label{sec:bccFe}

\subsubsection{Coupling constants for bcc Fe}
\label{sec:resJijPhiiijAijk}

In this Section we discuss the coupling parameters that we obtain from DFT, namely Heisenberg exchange $J_{ij}$, force constants $\phi_{ij}$, and exchange striction $\vec{A}_{ijk}$. 

The force constants are obtained from supercell calculations and are related by point-group symmetry  \cite{Ackland1997,Finkenstadt2006,Flocken1969,Frank1995}. Hence, we list only the irreducible values in Table \ref{tab:phiij}. In the table, $j = 1$ denotes the nearest neighbor shell, $j = 2$ the next nearest neighbor shell, and so forth. The indices $\mu$ and $\nu$ stand for the Cartesian coordinates.

\begin{table}[t]
\centering
\caption{
The nonvanishing elements $\phi^{\mu\nu}_{0j}$ of the force-constant matrix in Fe. The values are given in $\unit[]{mRyd \angstrom^{-2}}$.} 
{\begin{tabularx}{0.35\textwidth}{X X p{0.5cm} p{0.5cm} X r}
$j$ & & $\mu$ & $\nu$ & & $\phi_{0j}^{\mu \nu}$ ($\unit[]{mRyd \angstrom^{-2}}$)\\
\hline
\hline
  1 & & x        & x       & & -4360        \\
    & & x        & y       & & -2670         \\
\hline
  2 & & x        & x       & & -4850        \\
  2 & & y        & y       & & -2230        \\
\hline
  3 & & x        & x       & & -459        \\
    & & x        & y       & & -209        \\
    & & z        & z       & & 403         \\
\hline
  4 & & x        & x       & & 55        \\
    & & x        & y       & & -48         \\
    & & y        & y       & & -126         \\
    & & y        & z       & & -251         \\
\hline
  5 & & x        & x       & & 122         \\
    & & x        & y       & & -169        \\
\hline
\hline
 \end{tabularx}
 }
\label{tab:phiij}
\end{table}

\begin{figure}
\centering
\includegraphics[width=0.9\columnwidth]{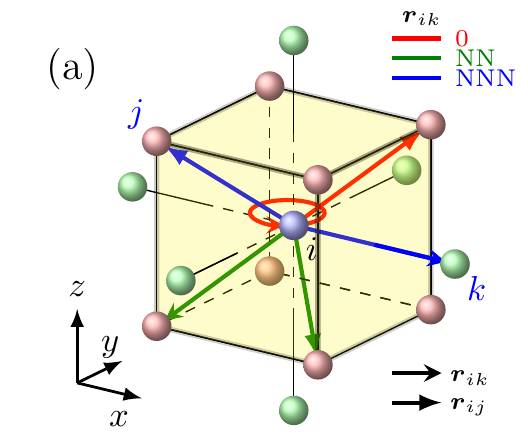}
\includegraphics[width=0.9\columnwidth]{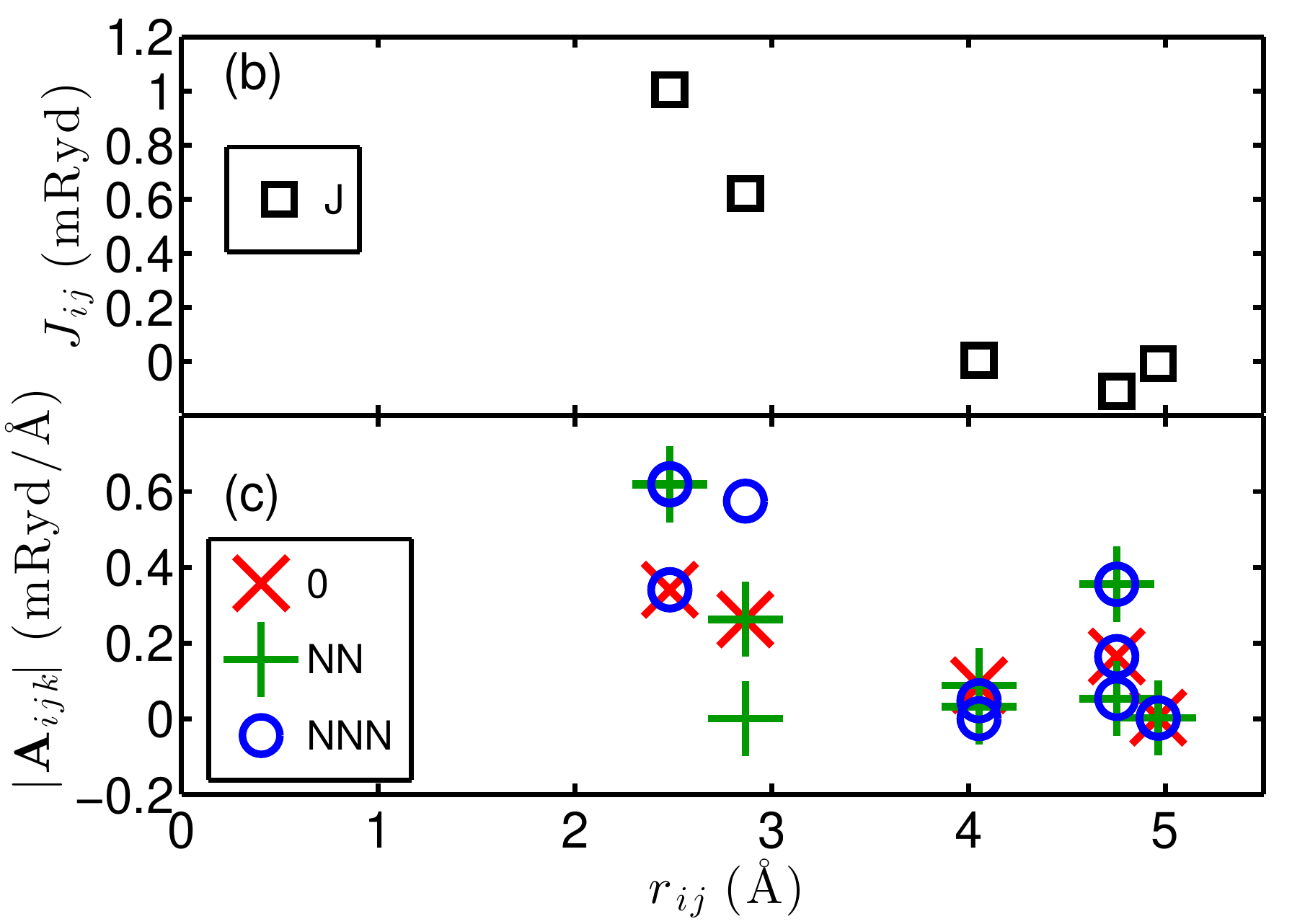}
\caption{\label{fig:aijk}(Color online) 
a) The nearest neighbor (NN) and next-nearest neighbor (NNN) coordination shells in bcc structure.
b) The Heisenberg exchange coupling $J_{ij}$ as a function of the distance $r_{ij}$.
c) Absolute value of the exchange striction $\vec{A}_{ijk}$ as a function of the distance $r_{ij}$, for on-site contribution (red cross), NN contribution (green plus), and NNN contribution (blue circle).
}
\end{figure}

The Heisenberg exchange for bcc Fe is well studied and our calculations agree well with previous studies  \cite{Pajda2001,Kvashnin2015,Szilva2017,Cardias2017}. The $J$'s are isotropic, long ranged, oscillating with decay typically as $r^{-3}$. The calculations reveal that $J\approx 0$ at $\unit[10]{nm}$. In Figure~\ref{fig:aijk})(b) the exchange is plotted against distance $r_{ij}$. The nearest neighbor interaction is about $\unit[1]{mRyd}$ and with also the second nearest neighbor interaction positive, bcc Fe is ferromagnetic. For the simulations in Sections \ref{sec:phasediag} and \ref{sec:sqwres} we use the first three coordination shells of exchange coupling.

From our first-principles calculations we obtained certain magnitude $\left|\vec{A}\right|=A$ and direction $\vec{e}^A$ of the spin-lattice coupling. We distinguish between the set of couplings $\{\vec{A}\}$ in a subset where the site $k$ is equal to site $i$ or $j$, e.g. $\vec{A}_{iji}$, and where $k\neq i,j$. The first subset refers to the couplings between two magnetic moments, where the lattice displacement happen on one of these two magnetic sites. Note that only these couplings were used, e.g. in Ref. \onlinecite{Ma2017}, although the many body property solved by density functional theory emphasize couplings at $k$ where $k\neq i,j$. 

The DFT calculations of $A_{iji}$ (red dots, Fig.~\ref{fig:aijk}) reveal a nearest-neighbor (NN) coupling of $A_{\mathrm{NN}}=\unit[0.34]{mRyd \angstrom^{-1}}$, where next-nearest neighbor (NNN) couplings are $A_{\mathrm{NNN}}=\unit[-0.26]{mRyd \angstrom^{-1}}$. Both align along the distance vector $\vec{r}_{ij}$. This implies that for this set of couplings the Heisenberg exchange $J_{ij}$ can be assumed to be a function only of the distance $\vec{r}_{ij}$. 
Here, $\nabla_i J_{ij}(\vec{r}_{ij})=\nicefrac{\partial J_{ij}}{\partial r_{ij}} \vec{e}^r_{ij}$. It also implies that the magnitude of the $\vec{A}_{iji}$ coupling is isotropic, and that the sum rule in Eq.~\ref{eq:sumrule} simplifies to $\vec{A}_{iji}=-\vec{A}_{ijj}$.

The strength of the exchange striction coupling differs in different direction. For instance, for the nearest and next-nearest neighbor coupling in both $\vec{r}_{ij}$ and $\vec{r}_{ik}$, the coupling $\vec{A}_{ijk}$ are bigger for the cases where $k\neq j,-j$. These couplings are $A_{\mathrm{NN,NN}}=\unit[0.62]{mRyd \angstrom^{-1}}$, $A_{\mathrm{NNN,NN}}=\unit[0.57]{mRyd \angstrom^{-1}}$, and $A_{\mathrm{NNN,NN}}=\unit[0.62]{mRyd \angstrom^{-1}}$. The $\vec{A}_{\mathrm{NNN,NNN}}$ couplings are only different from zero for the cases $k=j$.

\subsubsection{Thermodynamic properties of bcc Fe}
\label{sec:phasediag}
\begin{figure}
\includegraphics[width=0.50\textwidth]{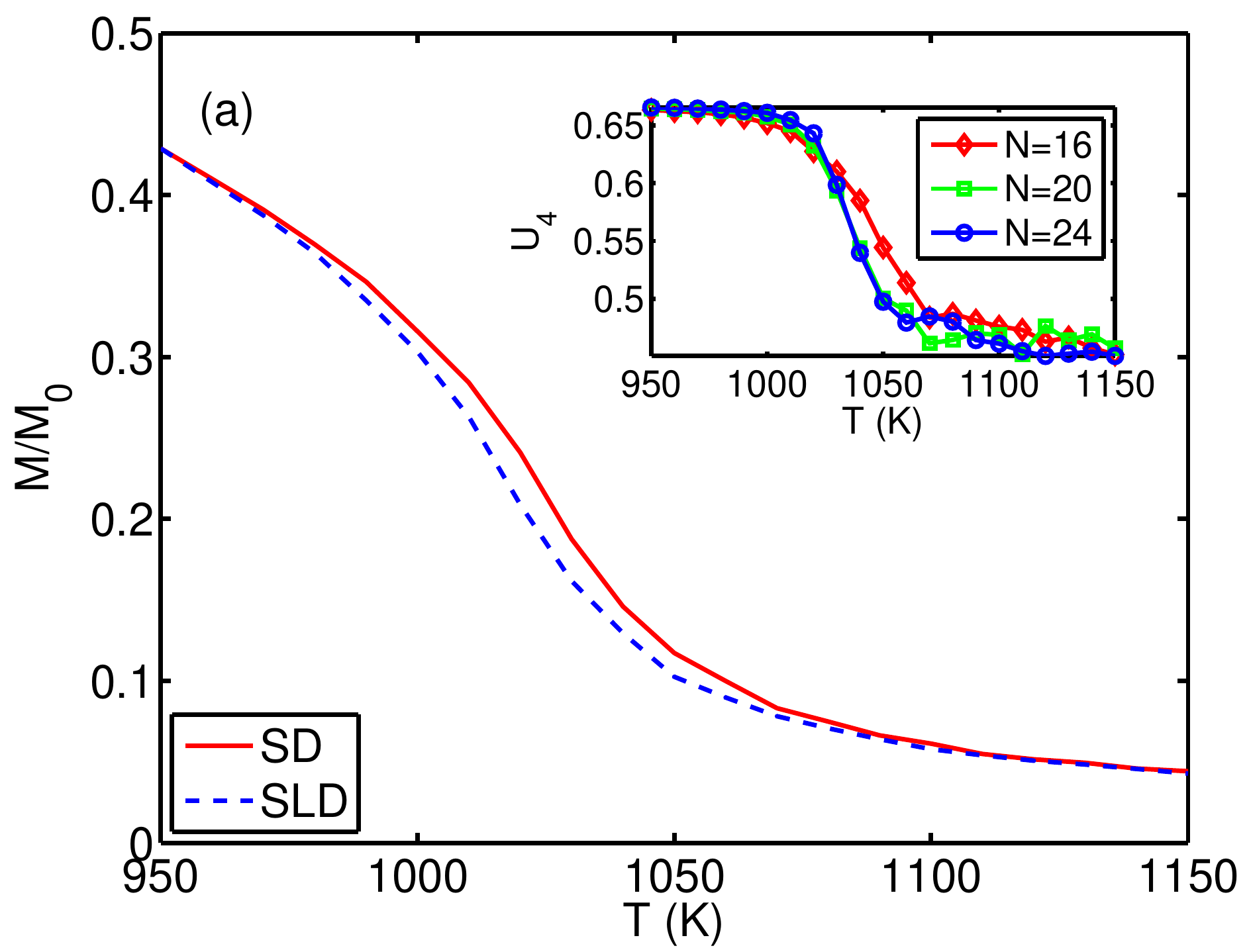}
\includegraphics[width=0.50\textwidth]{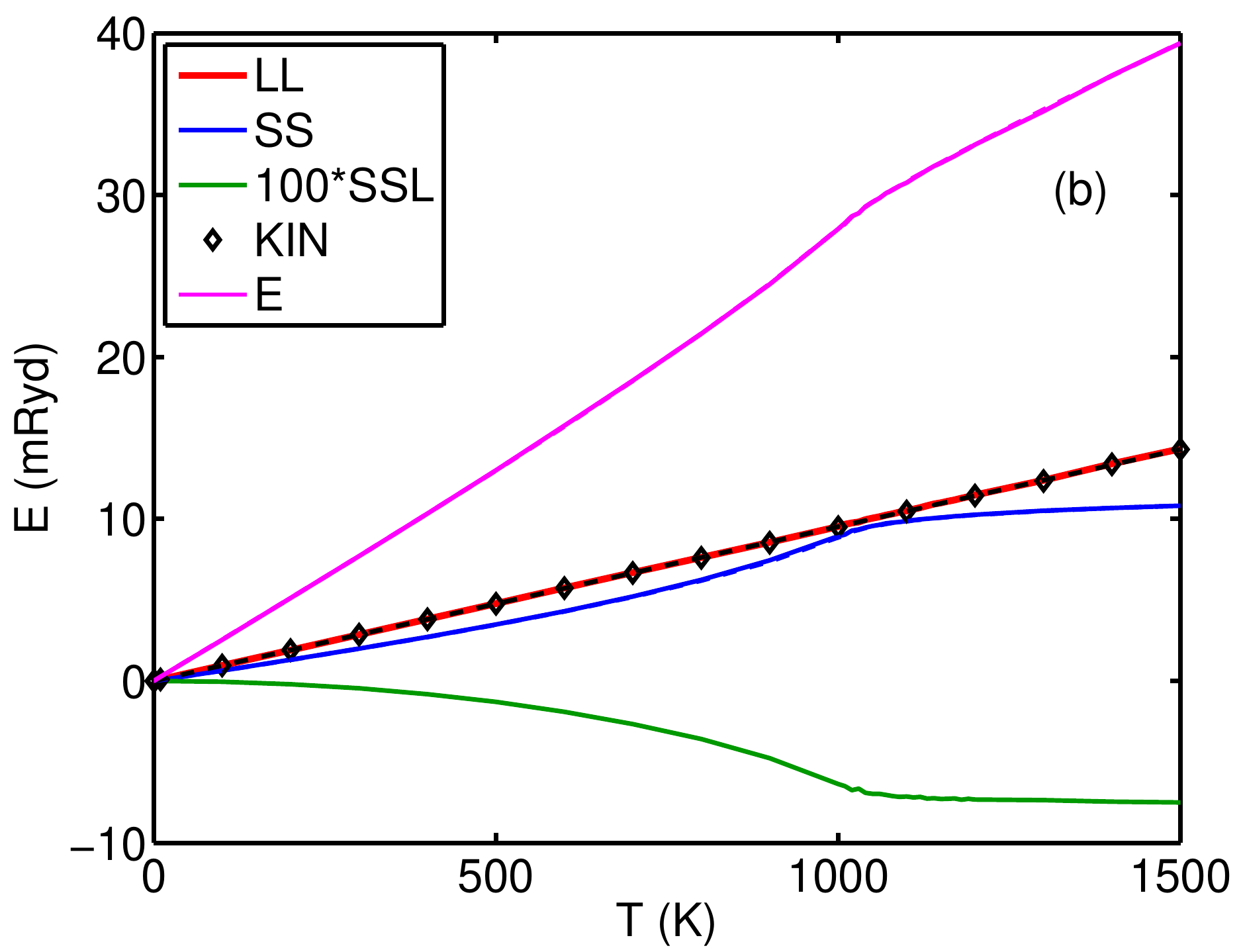}
\caption{\label{fig:bccfeN20} (Color online) (a) The magnetic order parameter M(T)/M(T=0) sampled in SLD and in SD simulation for a $N=20$ simulation cell. 
The inset shows the crossing of the $U_4$ Binder cumulant for cell sizes $N=16,20$ and 24 uncoupled SD simulations. 
(b) Energies of the spin-lattice dynamics Hamiltonian as a function of temperature.}
\end{figure}

In order to investigate how the presence of exchange striction affect the magnetization order parameter at finite temperature we have performed Langevin dynamics spin-lattice dynamics (SLD), and uncoupled spin-dynamics (SD) and lattice dynamics (LD), simulations in the temperature range 0 to 1500 K for simulation cells with size $N\times N\times N$ and periodic boundary conditions. 
For low and high temperatures the magnetic order parameters deviate negligibly when comparing SLD and SD data but, as shown in Fig. \ref{fig:bccfeN20}(a), in the vicinity of the phase transition temperature the order parameter takes a lower value in the SLD simulation than in the SD simulation. The inset shows the crossing of the $U_4$ Binder cumulant for cell sizes $N=16,20$ and 24 uncoupled LD/SD simulations from which a critical temperature $T_c=1020$ K can be read out. 
The energies of the different terms in the spin-lattice Hamiltonian are shown as a function of temperature in Fig. \ref{fig:bccfeN20}(b), and we can here observe that the difference in the energies for SLD simulation to energies for the uncoupled LD and SD simulation (not shown in figure), are smaller than the line-size in the graph, apart from, naturally, the exchange striction energy (SSL) which is identically zero in the uncoupled simulation and finite for the coupled system, this however changing the total energy (E) with a small fraction. The harmonic lattice potential (LL)  and the kinetic energy (KIN) coincide as expected given the equipartition theorem, and are linear in temperature. Unlike the lattice energies, the magnetic energy (SS) has an upper bound and flattens out above the phase transition temperature, and is taking a concave shape with values smaller than the lattice energy at lower temperature.

\begin{figure}
\includegraphics[width=0.50\textwidth]{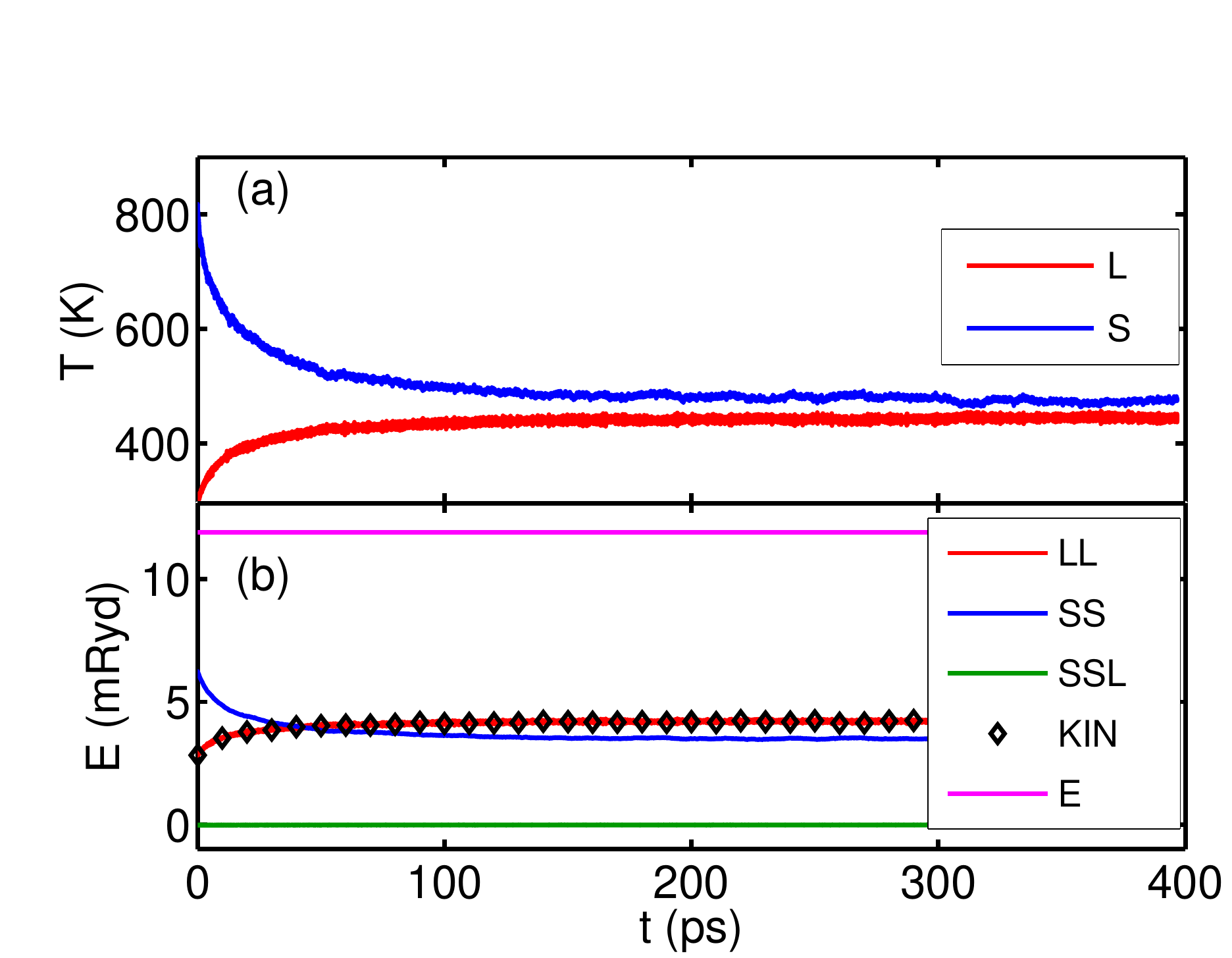}
\includegraphics[width=0.50\textwidth]{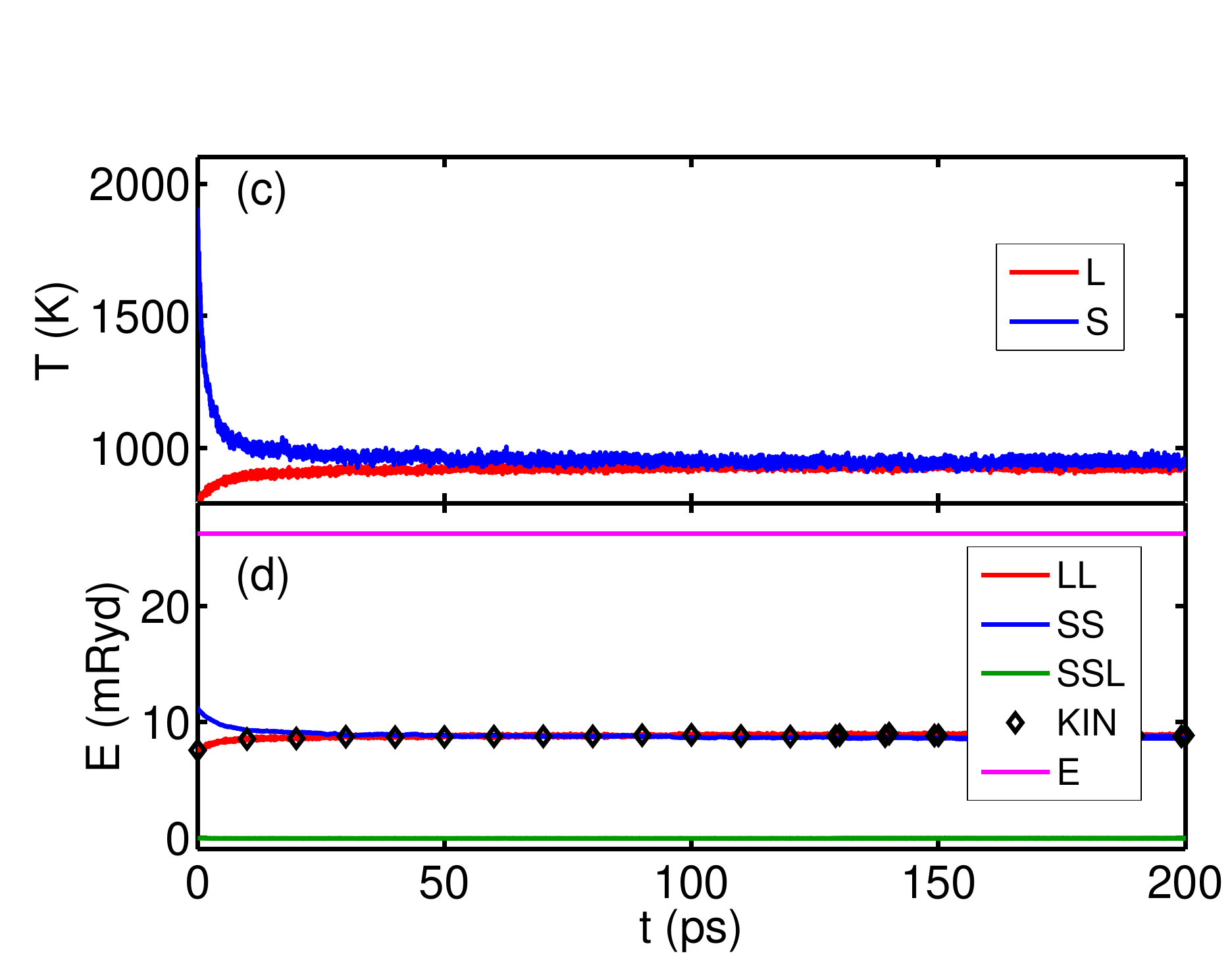}
\caption{\label{fig:relaxTemp} (Color online) 
Exchange striction mediated relaxation of the temperature and energy of spin and lattice subsystem in microcanonical evolution for two sets of initial temperatures of the subsystems. In a) and b) the initial spin temperature was 800 K and the initial lattice temperature was 300 K. In c) and d) the initial spin temperature was 2000 K and the initial lattice temperature was 800 K. At $t=0$ ps, the spin system and lattice system are each in thermal equilibrium with these heat baths. At $t>0$, the system evolves in Hamiltonian dynamics, with the total energy being conserved but being redistributed between the degrees of freedom.}
\end{figure}

In Figure \ref{fig:relaxTemp} is shown exchange striction mediated relaxation of the temperature and energy of spin and lattice subsystem in microcanonical evolution of the $N_{\rm atoms}=8000$ cell with edge length $N=20$, for undamped simulations. At $t=0$ ps, the spin system and lattice system are in thermal equilibrium with heat baths at different temperature. At $t>0$ the system evolve in Hamiltonian dynamics simulated with the fixed-point scheme for implicit mid-point method (see Appendix \ref{append:numint}) using a time step $dt=10^{-16}$ s. The total energy is conserved but is redistributed between the degrees of freedom. For initial conditions $T_S=800$ K and $T_L=300$ K are shown the time trajectories of temperature in \ref{fig:relaxTemp}(a) and energies in \ref{fig:relaxTemp}(b). As expected from the equipartition theorem, the lattice harmonic potential energy and kinetic energy take the same values. Complete equilibration of the spin and lattice subsystem to the same temperature does not happen during the displayed time interval of 400 ps. We attribute this incomplete relaxation to a kinematic constraint due to not only total energy, but also the total spin angular momentum, being constants of motion. At $t=0$ the average magnetization is $M\approx 1.3$ $\mu_B$ and as no net torque act on the spin system, energy can be transferred from the spin system to the lattice only in dynamics in which the magnetization is preserved. The situation is different in Figs. \ref{fig:relaxTemp}(c) and \ref{fig:relaxTemp}(d) where the initial condition is $T_S=2000$ K and $T_L=800$ K, and relaxation to a common temperature $T_S=T_L\approx 1000$ K occur, a process which is possible given that the system is paramagnetic at $T_S=2000$ K, and that the net equilibrium magnetization is very small at $T_S=1000$ K. At this temperature the harmonic potential, kinetic energy, and Heisenberg exchange coincide, in agreement with the thermal equilibrium data in Fig. \ref{fig:bccfeN20}. Finally, we note that in contrast to the presently discussed results for a ferromagnetic system, antiferromagnetic dynamics allow for relaxation without an angular momentum bottleneck, see, e.g., the relaxation in Hamiltonian dynamics reported in Ref.~\onlinecite{Hellsvik2016}.

\subsubsection{bcc Fe Magnon and phonon dispersions}
\label{sec:sqwres}

The simulations that we have performed to sample the dynamic structure factor for bcc Fe were performed in two stages: 
i) Equilibration stage with Langevin dynamics simulated with the combined velocity-Verlet and SIB solver algorithm, as described in the Appendix~\ref{append:numint}, subdivided into phases with first a longer time step ($dt=10^{-15}$~s) and high damping, followed by gradually shorter time steps and lower damping. In the fourth and final phase we used $N_t=10^4$ time steps of $dt=10^{-16}$~s and the damping parameters $\alpha=0.01$ and $\nu=10^{-14}$ kg$/$s, 
ii) Measurement stage done in Hamiltonian evolution of the system over $N_t=2\cdot 10^5$ time steps of $dt=5\cdot10^{-16}$~s with the fixed-point iteration implicit midpoint method, see Appendix~\ref{append:numint}. The sampling step $t_{\rm samp}=5\cdot10^{-15}$~s for the correlation functions defined in Eqs.~\ref{eq:cmrt}-\ref{eq:cvrt} is used for a sampling window of $t_{\rm win}=5\cdot10^{-11}$~s and combined with averaging of the correlations by moving the time window over $t_0=(0,5,10,\ldots,5\cdot10^4)10^{-15}$~s. The corresponding frequency range for the dynamic structure factors in Eqs.~\ref{eq:msqw}-\ref{eq:vsqw} is $\nicefrac{\omega}{2\pi}=[0.02, 0.04,\ldots,200]$~THz (0.0827 meV to 827 meV). 

\begin{figure}
\includegraphics[width=0.50\textwidth]{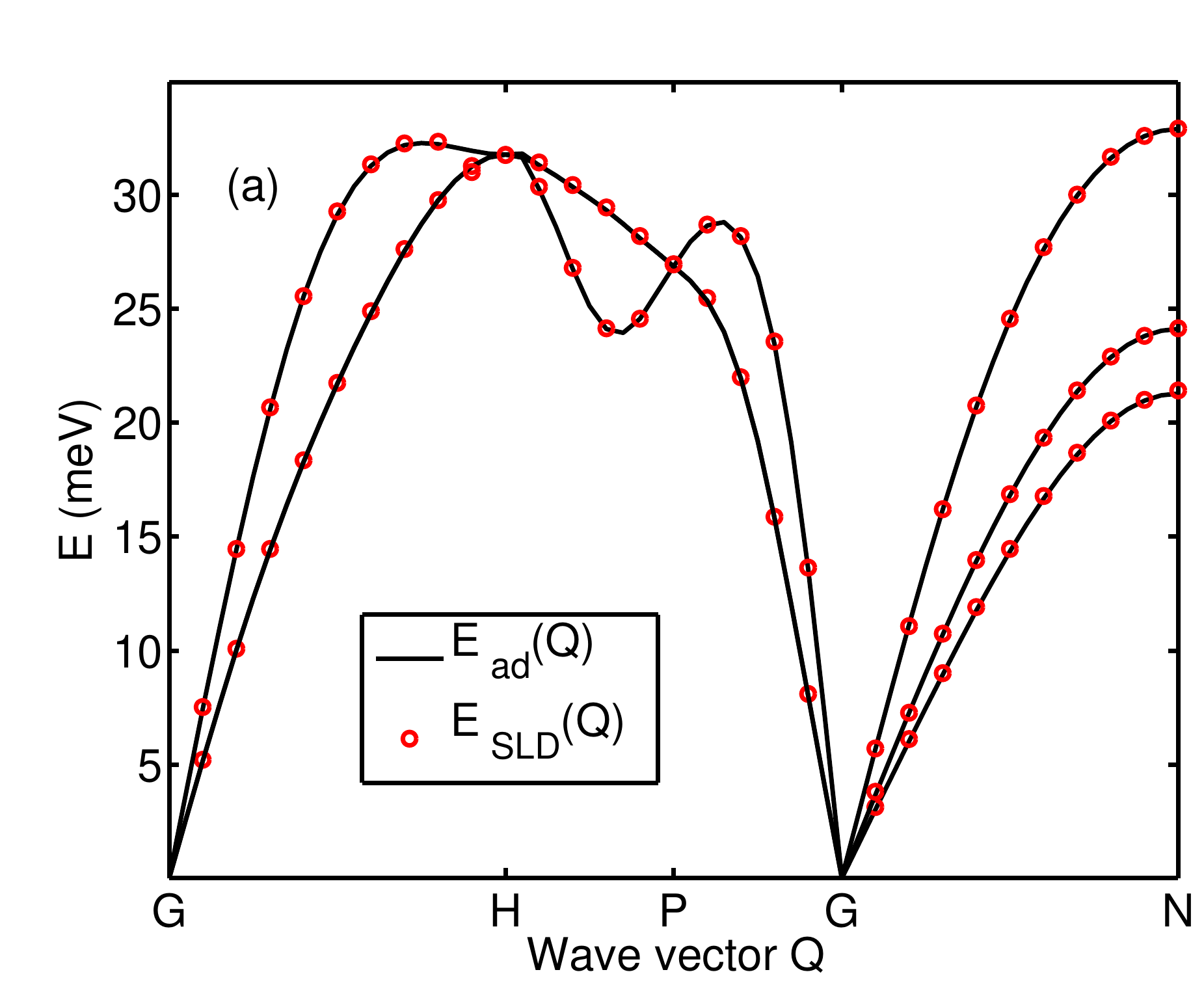}
\includegraphics[width=0.50\textwidth]{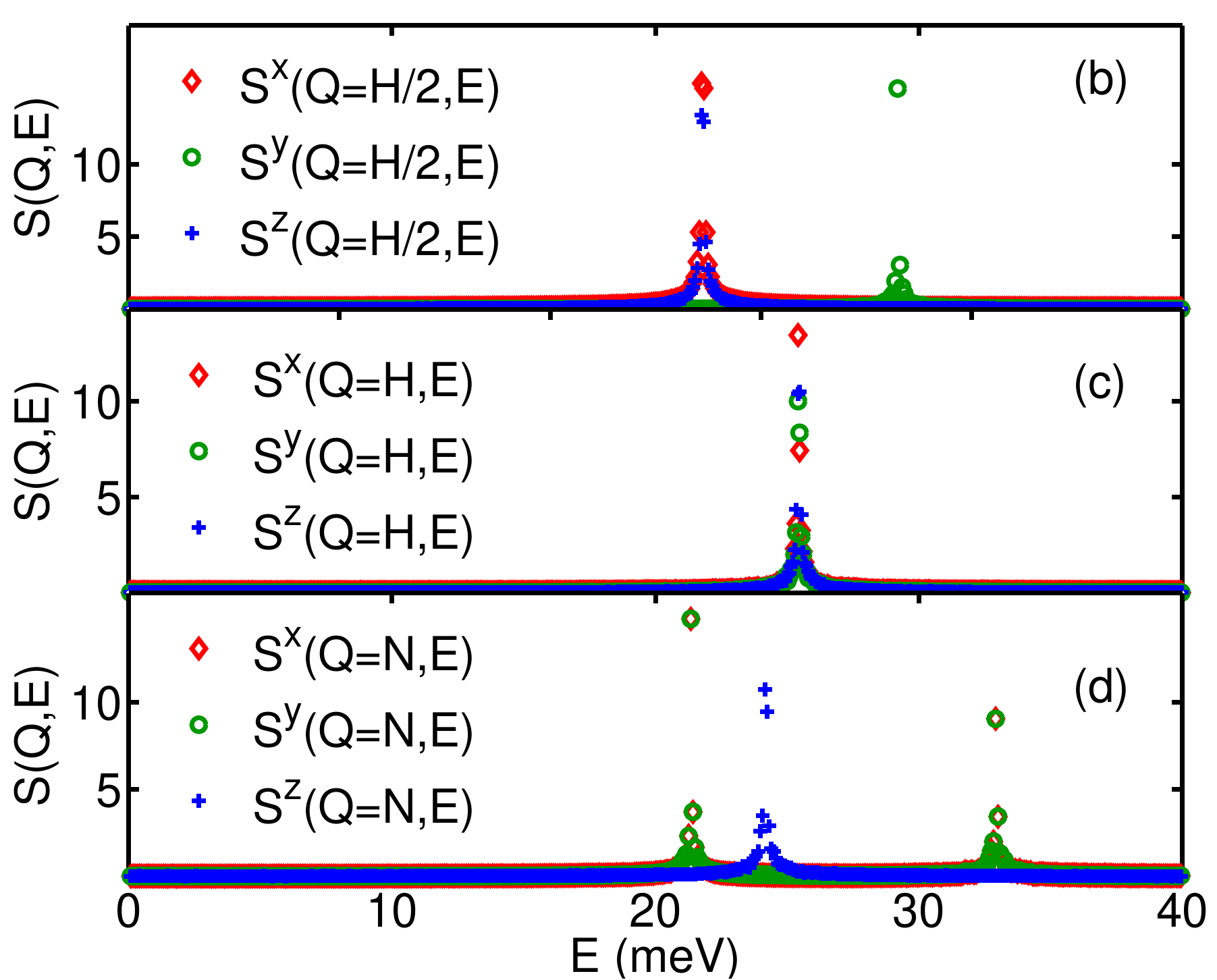}
\caption{\label{fig:sqwuSLDT300} 
 (Color online) 
(a) The $T=0$ K adiabatic phonon dispersion $\omega(\bfqq)$ (black curve) calculated using Eq. \ref{eq:phondisp}, and the peaks of the 
displacement-displacement dynamic structure factor $S(\bfqq,E)$ (red symbols) sampled in SLD simulations of bcc Fe at $T=300$~K using a 20$\times$20$\times$20 supercell with periodic boundary conditions. 
The dynamic structure factor for selected q-points (b) 
H/2 , (c) H and (d) N points.
}
\end{figure}

In order to investigate the impact of the exchange striction on the magnon and phonon spectra at finite temperature, we pursued both spin-lattice dynamics simulations and uncoupled spin dynamics and lattice dynamics simulations. In Figure \ref{fig:sqwuSLDT300} is shown the displacement-displacement dynamic structure factor $S(\bfqq,E)$ sampled in SLD simulation at $T=300$ K, as well as the $T=0$ K adiabatic phonon dispersion $\omega(\bfqq)$ obtained from use of Eq. \ref{eq:phondisp}.

\begin{figure}
\includegraphics[width=0.50\textwidth]{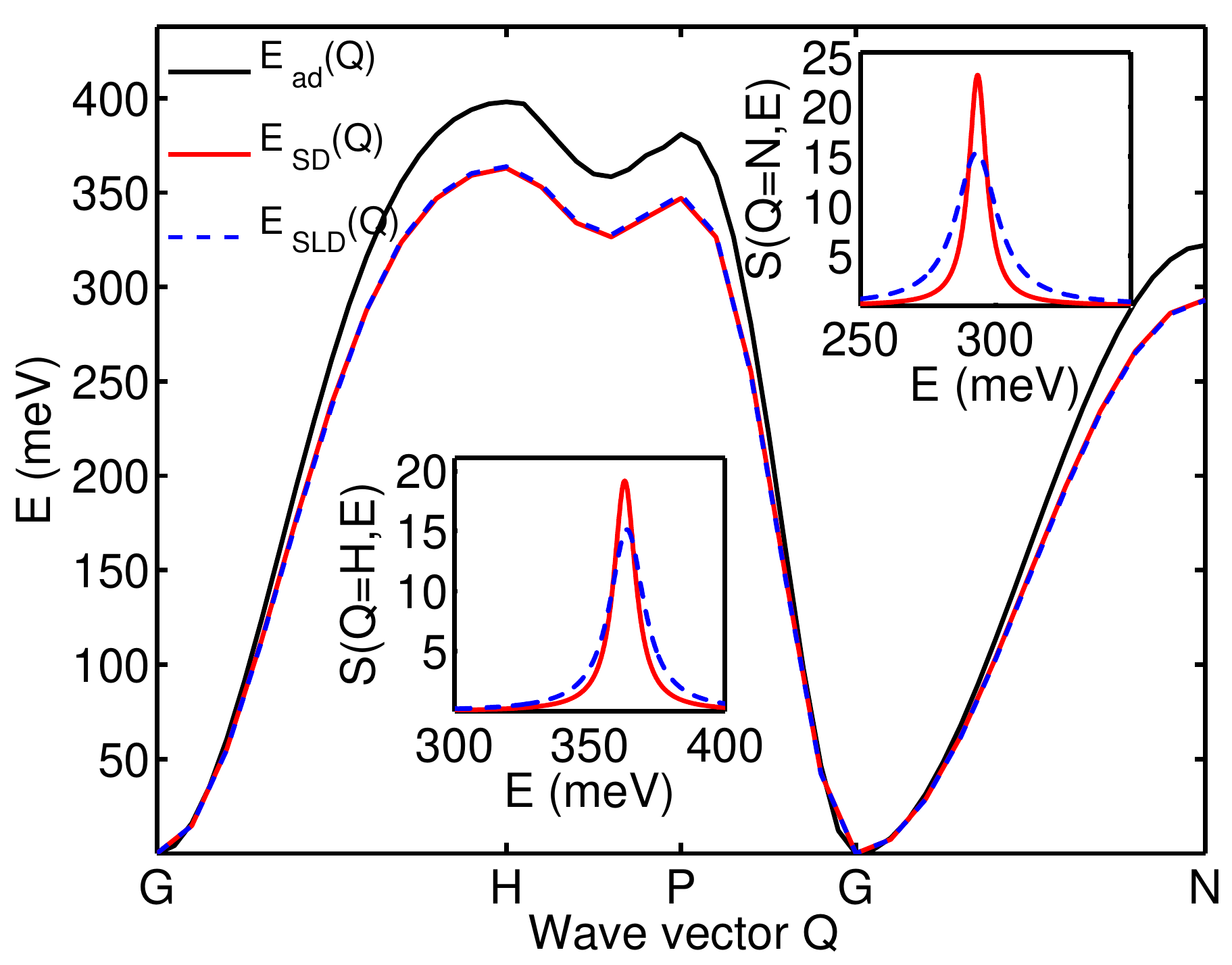}
\caption{\label{fig:sqwspindiffT300} 
(Color online) The $T=0$ K adiabatic magnon dispersion $\omega(\bfqq)$ (black) calculated using Eq. \ref{eq:AMS}, and the peak of the $x$-component of the spin-spin dynamic structure factor $S(\bfqq,E)$ sampled in SLD (blue) and SD (red) simulations of bcc Fe at $T=300$~K using a 20$\times$20$\times$20 supercell with periodic boundary conditions. The insets shows the shape of the Lorentzian functions at the H and N points for SLD and SD simulation respectively.}
\end{figure}

Similarly, in Figure \ref{fig:sqwspindiffT300} we display the $x$-component of the spin-spin dynamic structure factor $S(\bfqq,E)$ and the $T=0$ K adiabatic magnon dispersion $\omega(\bfqq)$ (black) calculated using Eq. \ref{eq:AMS}. The insets show the shape of the Lorentzian functions at the H and N points for SLD and SD simulation respectively. The presence of exchange striction in the SLD simulation causes a broadening of the resonance peaks as compared to the SD simulation.

\begin{figure}
\includegraphics[width=0.50\textwidth]{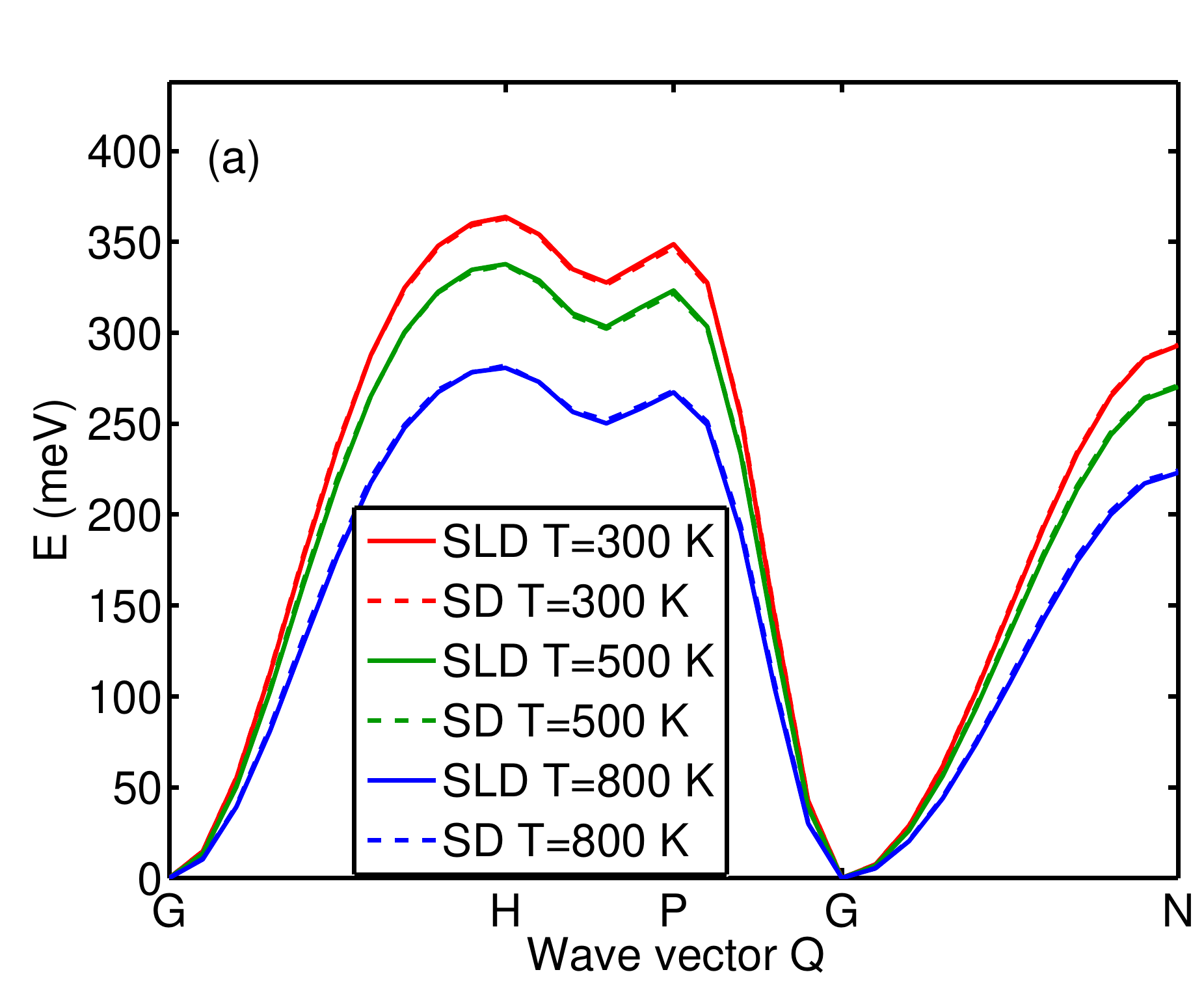}
\includegraphics[width=0.50\textwidth]{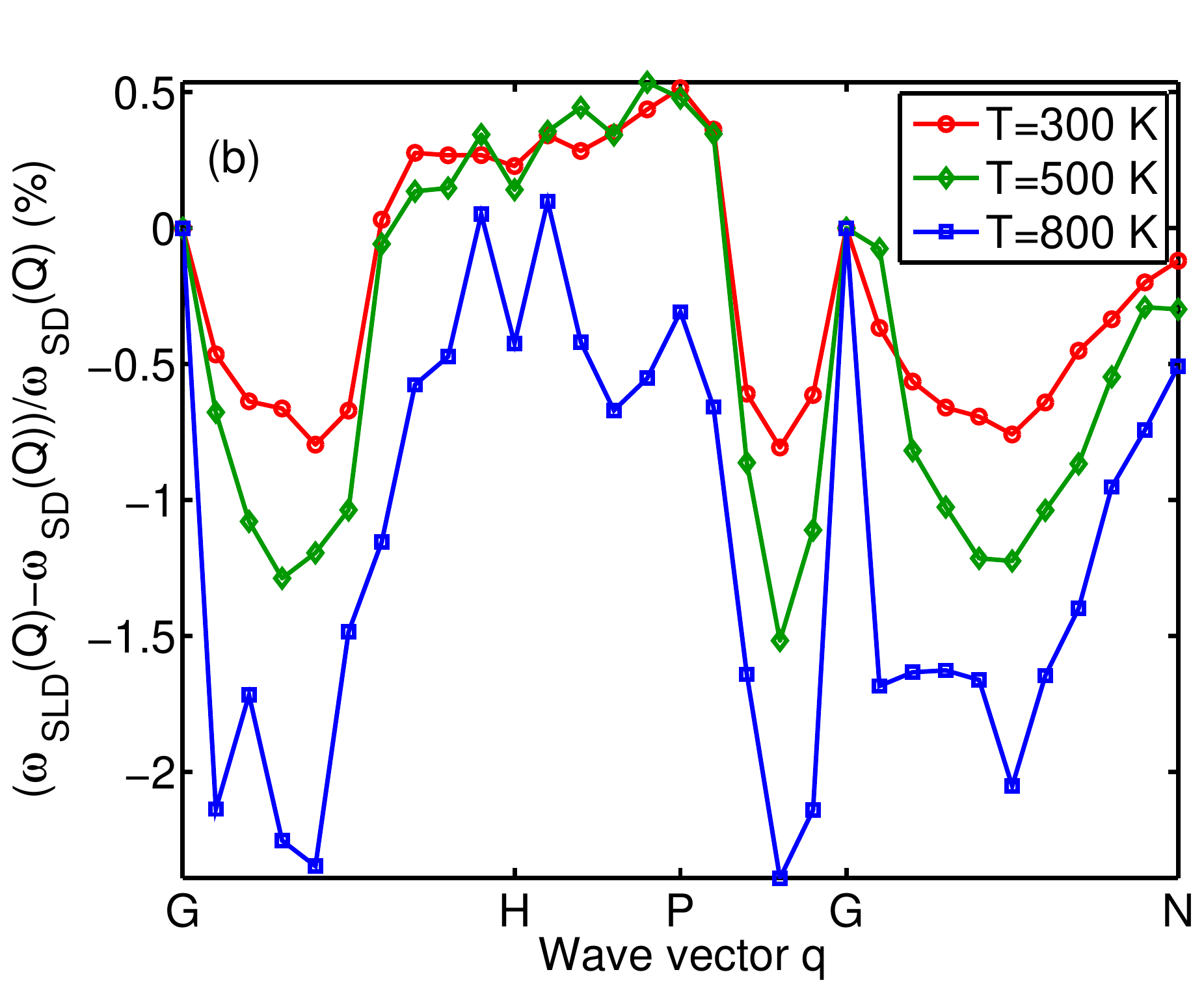}
\caption{\label{fig:bccfeSQW} 
(Color online) The dispersions fitted from the spin-spin dynamic structure factor $S(\bfqq,E)$ sampled in SLD simulations and in SD simulations and lattice dynamics simulations of bcc Fe at different temperatures using a 20$\times$20$\times$20 supercell with periodic boundary conditions. 
(a) The finite temperature dispersion $\omega(\bfqq)$ fitted with a Lorentzian function from $S(\bfqq,E)$ SLD (full lines) and SD (dashed lines) simulations at $T=300$ K (red), $T=500$ K (green), and $T=800$ K (blue).
(b) The ratio $(\omega(\bfqq)_{SLD}-\omega(\bfqq)_{SD})/\omega(\bfqq)_{SD}$, at $T=300$ K (red), $T=500$ K (green), and $T=800$ K (blue).}
\end{figure}

Shown in the upper panel of Fig. \ref{fig:bccfeSQW} are the peak positions of the magnon dispersions at different temperature, obtained by means of fitting to a Lorentzian function. In the very detailed investigations by Perera \emph{et al.} \cite{Perera2017} of magnon and phonon spectra of the Dudarev-Derlet potential  \cite{Dudarev2005,Derlet2007} potential for bcc Fe, the measure $(\omega_{\rm SLD}(Q)-\omega_{\rm SD}(Q))/\omega_{\rm SD}(Q)$ was used to analyze the temperature-dependent influence of exchange striction on magnon dispersion. Similarly, the quantity $(\omega_{\rm SLD}(Q)-\omega_{\rm LD}(Q))/\omega_{\rm LD}(Q)$ was defined for the phonon dispersions. In the lower panel of Fig. \ref{fig:bccfeSQW} we show results for the ratio $(\omega_{\rm SLD}(Q)-\omega_{\rm SD}(Q))/\omega_{\rm SD}(Q)$ and note that our results for a Hamiltonian constructed by means of first principles density functional theory methods, compare well with the results obtained by Perera \emph{et al.} for the Dudarev-Derlet potential.

Overall our results compare well with \cite{Brockhouse1967,Minkiewicz1967,Klotz2000} (experimental) and theoretical \cite{Muller2007,Perera2017} (theory) phonon dispersions at finite temperature, and with \cite{Lynn1975} (experiment) and \cite{Kvashnin2015} (theory) magnon dispersions.

\section{Conclusion}
\label{sec:conc}

In conclusion, we have demonstrated a computationally efficient general method for performing spin-lattice coupled simulations. The method is, in short, based on a Taylor expansion of the bilinear magnetic term of the Hamiltonian with respect to motion of atmic moments and nuclear positions.
To test the reliability of our method, we checked it against available analytical results, obtaining excellent agreement. In our conceptual simulations for small magnetic clusters, we observe new modes emerging as a result of strong interaction between atomic and spin motion. We propose that these coupled modes should be detectable in Raman spectroscopy. We also performed simulations of bulk bcc iron obtaining very good agreement with previous simulations based on an empirical Hamiltonian.
In general, the interaction between the spin and lattice degrees of freedom can be expected to lead to significant changes in both the magnon and phonon spectra, and our simulations indeed demonstrate this. Also, as expected, the changes tend to become more pronounced as the temperature is increased. 

\section{Acknowledgements}
We acknowledge fruitful discussions with Pavel F. Bessarab, Jonas Fransson, Pablo Maldonado, Johan H. Mentink, and Lars Nordström. 
Financial support from Vetenskapsrådet (grant numbers VR 2015-04608 and VR 2016-05980), and Swedish Energy Agency (grant number STEM P40147-1) is acknowledged. J.H. is partly funded by the Swedish Research Council (VR) through a neutron project grant (BIFROST, Dnr. 2016-06955). O.E. acknowledges the support from  Swedish Research Council (VR), eSSENCE, the Knut and Alice Wallenberg (KAW) Foundation (Grants No. 2012.0031 and No. 2013.0020), STandUPP, and the foundation for strategic research (SSF). L.B. acknowledges the support from Swedish Research Council (VR), Grant. No. 2017-03763. 
The computations were performed on resources provided by the Swedish National Infrastructure for Computing (SNIC) at the  Uppsala Multidisciplinary Center for Advanced Computational Science (UPPMAX), the National Supercomputer Center (NSC), Linköping University, the PDC Centre for High Performance Computing (PDC-HPC), KTH, and the High Performance Computing Center North  (HPC2N), Umeå University.

\appendix

\section{Numerical integration of the spin-lattice dynamics equations of motion}
\label{append:numint}

In this appendix, the schemes we use for numerical integration of the coupled equation of motions expressed in Eqs.~(\ref{eq:SLDeom1}-\ref{eq:SLDeom3}) are described.
Explicit methods for integrating the stochastic LLG equation are commonly two-step numerical integration as is the case for the Heun method~\cite{Rumelin1982}, the Depondt-Merten's method~\cite{Depondt2009}, and the semi-implicit SIB method by Mentink \emph{et al.}~\cite{Mentink2010}. A description of these methods, including benchmarks, can be found in Ref.~\cite{Eriksson2017asd}.
The Depondt-Merten's method and the semi-implicit SIB method can be extended with a suitable explicit or semi-implicit solver for the lattice degrees of freedom, such as the velocity-Verlet method. Note that integration with Heun or other explicit Runge-Kutta schemes is well known to have poor stability for molecular dynamics.

For the Hamiltonian simulations we use a fixed-point iteration of the implicit midpoint scheme. For the simulations in the canonical ensemble, we use a combination of the Gr{\o}nbech-Jensen and Farago (G-JF) \cite{Gronbech-Jensen2013} Verlet-type methods for simulation of Langevin molecular dynamics and the Mentink \emph{et al.} semi-implicit SIB method for the stochastic LLG equation~\cite{Mentink2010}. The combined algorithm for the canonical simulation is written out in pseudocode below.
\\

\begin{algorithmic}
\For {$k \leq K$}	\Comment{Loop over time step}\\
	\For {$i \in N_{\rm mag}$}	\Comment{The first step}
		\State calculate $\bfbb_k^i (\bfu_k,\bfm_k)$
	\EndFor
	\For {$i \in N_{\rm all}$}
		\State calculate $\bfff_k^i (\bfu_k,\bfm_k)$
	\EndFor
	\For {$i \in N_{\rm mag}$}
		\State calculate $\tilde{\bfm}_{k+1}^i (\bfm_k^i,\tilde{\bfm}_{k+1}^i,\bfbb_k^i)$   \Comment{Implicit in $\tilde{\bfm}_{k+1}^i$}
	\EndFor
	\For {$i \in N_{\rm all}$}
		\State calculate $\bfu_{k+1}^i (\bfu_k^i,\bfv_k^i,\bfff_k^i)$   \Comment{One-shot}
	\EndFor\\
	\For {$i \in N_{\rm mag}$}	\Comment{The second step}
		\State calculate $\bfbb_{k+1}^i (\bfu_{k+1},\tilde{\bfm}_{k+1})$
	\EndFor
	\For {$i \in N_{\rm mag}$}
		\State calculate $\bfm_{k+1}^i (\bfm_k^i,\bfbb_k^i,\bfm_{k+1}^i,\bfbb_{k+1}^i)$   \Comment{Implicit in $\bfm_{k+1}^i$}
	\EndFor
	\For {$i \in N_{\rm all}$}
		\State calculate $\bfff_{k+1}^i (\bfu_{k+1},\bfm_{k+1})$
	\EndFor
	\For {$i \in N_{\rm all}$}
		\State calculate $\bfv_{k+1}^i(\bfv_k^i,\bfff_k^i,\bfff_{k+1}^i)$   \Comment{One-shot}
	\EndFor\\
\EndFor
\end{algorithmic}

\end{document}